\documentclass[10pt, twoside, twocolumn]{IEEEtran}
 \makeatletter

\newcommand{\Rmnum}[1]{\expandafter\@slowromancap\romannumeral #1@}
\makeatother
\usepackage{graphicx,cite,epsfig,amssymb,amsmath,multirow,cases}
\usepackage{amsfonts}
\usepackage{mathrsfs}
\usepackage{algorithm}
\usepackage{algorithmic}
\usepackage{float}
\usepackage{bm}
\usepackage{booktabs}
\usepackage{epstopdf}
\usepackage{ulem}
\usepackage{slashbox}
\usepackage[usenames,dvipsnames]{color}
\usepackage{subfigure}
\usepackage{color}

\newtheorem{lemma}{Lemma}
\makeatother
\begin{document}

\bibliographystyle{ieeetr}
\title{Millimeter-Wave Massive MIMO Systems Relying on Generalized Sub-Array-Connected Hybrid Precoding}
\author{Yun~Chen,~Da~Chen,~Tao~Jiang, \emph{Fellow, IEEE},~and~Lajos~Hanzo, \emph{Fellow, IEEE}
\thanks{Manuscript received November 24, 2018; revised April 26, 2019; accepted July 16, 2019. This work was supported in part by the National Science Foundation of China with Grant numbers 61771216 and 61631015, Fundamental Research Funds for the Central Universities with Grant number 2015ZDTD012, and China Scholarship Council (CSC). L. Hanzo would like to acknowledge the financial support of the Engineering and Physical Sciences Research Council projects EP/Noo4558/1, EP/PO34284/1, COALESCE, of the Royal Society's Global Challenges Research Fund Grant as well as of the European Research Council's Advanced Fellow Grant QuantCom. (\emph{Corresponding author: Lajos Hanzo.})

Y.~Chen,~D.~Chen,~and~T.~Jiang are with Wuhan National Laboratory for Optoelectronics, School of Electronic Information and Communications, Huazhong University of Science and Technology, Wuhan 430074, China (e-mail: chen\_yun@hust.edu.cn; chenda@hust.edu.cn; tao.jiang@ieee.org).

L. Hanzo is with School of Electronics and Computer Science, University of Southampton, Southampton SO17 1BJ, U.K. (e-mail: lh@ecs.soton.ac.uk).

Copyright (c) 2015 IEEE. Personal use of this material is permitted. However, permission to use this material for any other purposes must be obtained from the IEEE by sending a request to pubs-permissions@ieee.org.
}}

\markboth{IEEE Transactions on Vehicular Technology ,~Vol.~, No.~, 2019}%
{Chen \MakeLowercase{\textit{et al.}}: Generalized Sub-Array-Connected Hybrid Precoding}

\maketitle
\begin{abstract}
In this paper, we consider a generalized sub-array-connected (GSAC) architecture for arbitrary radio frequency (RF) chain and antenna configurations, where the number of RF chains connected to a sub-array and the number of antennas in each sub-array can be arbitrary. Our design objective is to improve the energy-efficiency of the hybrid precoder of millimeter-wave massive multiple input multiple output (MIMO) systems. We firstly propose a successive interference cancellation based hybrid precoding algorithm to maximize the achievable rate for any given RF chain and antenna configuration. This algorithm firstly decomposes the total achievable rate optimization problem into multiple sub-rate optimization problems, then it successively maximizes these sub-rates. Since the number of RF chains is limited, we can afford using an exhaustive search scheme to evaluate all configurations and identify the one having the best energy efficiency. Moreover, to rely on an attractive limited feedback, we also propose a beamsteering codebook for our hybrid precoding aided GSAC architecture. Our simulation results demonstrate that the proposed scheme achieves a similar rate as the corresponding optimal unconstrained precoder. Furthermore, we show that the energy-efficiency of the proposed scheme is better than that of the existing schemes in the fully-connected and sub-array-connected architectures.
\end{abstract}


%\nomenclature[Cp]{$p_{Di}$}{Active power demand at bus $i$}
%\nomenclature[Vp]{$p_{Gi}$}{Active power generation at bus $i$}
%\nomenclature[SO]{$\Omega_{G}$}{Set of generator buses}

\section{Introduction}
Millimeter-wave massive multiple input multiple output (MIMO) systems constitute promising candidate technologies for next-generation communication systems as a benefit of their substantial bandwidth and high spectral efficiency \cite{Hemadeh2018}. For example, at 30 GHz carrier frequency, the wave-length is 10 millimeters, which makes it possible to pack a large number of antennas in a compact area. As a benefit, a large antenna array is capable of providing significant precoding gains to compensate for the high path loss of millimeter-wave signals \cite{Heath2016}. However, in traditional MIMO systems, the Transmit Prec-\\oding (TPC) is usually realized in the digital domain and re-\\quires the same number of radio frequency (RF) chains as the number of antennas {\cite{PLiu2017, SQiu2018, Ticao2016, Asaad2018, Bereyhi2018}}. Hence digital TPC potentially imposes prohibitive energy consumption in millimeter-wave massive MIMO systems relying on large antenna arrays \cite{7Jin2012, 8Jin2016, Mirza2018}. To circumvent this problem, the hybrid TPC concept has been proposed, where the signals are firstly precoded by a low-dimensional digital TPC to cancel the interference between data streams and to allocate the transmit power. Then they are also precoded by a high-dimensional analog TPC to attain high beamforming gains \cite{Ayach2014, Hanzo2018, YChenTCOM, YChenIOT}.

Most hybrid TPC schemes consider the fully-connected (FC) and the sub-array-connected (SAC) architectures \cite{14Rusu2016, 19Chen2017, 11Ni2016, Yu2016, 13Chen2015, Gao2016, Park2017}. In the FC architecture, each RF chain is connected to all antennas by a large number of analogue phase shifters to achieve the maximum attainable TPC gains, which however leads to a high energy consumption \cite{Heath2016}. The orthogonal matching pursuit (OMP) based hybrid TPC algorithm was the first one proposed for the FC architecture in millimeter wave MIMO systems, which formulates the spectral efficiency optimization as a sparse reconstruction problem \cite{Ayach2014}. Since the OMP based algorithm is of high computational complexity, numerous authors designed low-complexity hybrid TPC schemes \cite{14Rusu2016, 19Chen2017}. Moreover, the alternating minimization, matrix decomposition and iterative search based hybrid TPC schemes were also proposed for further improving the spectral efficiency \cite{11Ni2016, Yu2016, 13Chen2015}. By contrast, the SAC architecture requires a lower number of phase shifters, but has to tolerate some loss of the achievable rate \cite{Gao2018}. The authors of \cite{Gao2016} are the first to consider the design of hybrid TPC schemes for the SAC architecture, where successive interference cancellation (SIC) was involved separately for optimizing the achievable rate of each sub-array. In \cite{Yu2016}, the vector approximation and semi-definite programming (SDP) techniques were utilized to design a hybrid TPC scheme relying on the diagonal structure of the SAC architecture. In \cite{Park2017}, A closed-form solution was proposed for the SAC architecture in the context of a wide-band millimeter wave system.

There is a tradeoff between the achievable rate and the energy efficiency of the FC and SAC architectures. The best hybrid TPC architecture having the highest energy efficiency (EE) is still unknown at the time of writing. However, efforts have been invested into improving the EE of millimeter wave MIMO systems \cite{Roth2017, He2017, Moghadam2018}. But the above papers only considered the EE optimization under a particular hybrid TPC architecture and did not exploit the full potential of the hybrid TPC architecture itself. Recently, a more general SAC architecture, termed as hybrid-connection based architecture was proposed, where each sub-array may be connected to multiple RF chains \cite{Hanzo2018, Zhang2018}. The authors of \cite{Hanzo2018} further separated the different sub-arrays by a sufficiently large distance for achieving both TPC and diversity gains simultaneously. However, the hybrid-connection based architecture assumes that the number of RF chains for all sub-arrays is the same, which limits the degree of freedom in improving the EE. Against this background, our novel contributions are:

\begin{itemize}
\item
  We propose a generalized sub-array-connected (GSAC) architecture, where the number of RF chains connected to a sub-array and the number of antennas in each sub-array can be arbitrarily adjusted for improving the EE of hybrid TPC in millimeter-wave massive MIMO systems.

\item For any given RF and antenna configuration in the GSAC architecture, a SIC and phase extraction based hybrid TPC algorithm is proposed. This scheme firstly decomposes the total achievable rate optimization problem into multiple sub-rate optimization problems, each of which is only related to a single sub-array.
  Then, it successively maximizes these sub-rates.

\item Since the typical millimeter wave channel exhibits limited scattering and because the total number of RF chains is limited, we can afford using an exhaustive search scheme to determine the RF and antenna configuration of the GSAC architecture having the highest energy efficiency.

  \item Moreover, to be able to rely on an attractive limited feedback, where only the receiver has the channel state information (CSI), we also propose a beamsteering codebook based hybrid TPC scheme for our GSAC  architecture. Our simulation results demonstrate that the proposed scheme achieves a similar rate as the corresponding optimal unconstrained TPC scheme and the EE of the proposed scheme is the best in the family of the FC and SAC architectures.
\end{itemize}

The remainder of this paper is organized as follows. In Section II, the system model and the channel model are described. The GSAC architecture, the SIC-based hybrid TPC scheme designed for our GSAC architecture, and the exhaustive search scheme are discussed in Section III, while Section IV elaborates on our hybrid TPC design operating on limited feedback. Our simulation results are presented in Section V. Finally, we conclude this paper in Section VI.

\emph{Notation}: $ a$ and $A$ are scalars, $\bf{a}$ is a vector, and $\bf{A}$ is a matrix. ${\left\| {\bf{a}} \right\|_1}$ and ${\left\| {\bf{a}} \right\|_2}$ denote the $l_1$ and $l_2$ norm of $\bf{a}$, respectively. ${{\bf{A}}^T},{{\bf{A}}^*},{{\bf{A}}^{ - 1}}$, $\left| {\bf{A}} \right|$ and ${\left\| {\bf{A}} \right\|_F}$ denote the transpose, conjugate transpose, inverse, determinant, and Frobenius norm of ${\bf{A}}$, respectively. ${{\bf{I}}_N}$ denotes a $N\!\times\! N$ identity matrix. $\mathbb{E}[\cdot]$ denotes the expectation.

\section{System Model and Channel Model}
We consider a single-user millimeter-wave massive MIMO system, where the transmitter is equipped with $N_{\rm{t}}$ antennas and $N^{\rm t}_{\rm{RF}}$ RF chains.
$N_{\rm{s}}$ data streams are transmitted to the receiver having $N_{\rm{r}}$ antennas and $N^{\rm r}_{\rm{RF}}$ RF chains. The $N_{\rm{r}} \times 1$ received signal vector $\bf y$ can be presented as
\begin{equation}\label{2.2}
{\bf{y}} = \sqrt \rho  {\bf{H}}{{\textbf{F}}_{{\rm{RF}}}}{{\textbf{F}}_{{\rm{BB}}}}{\bf{s}} + {\bf{n}},
\end{equation}
where $\rho$ is the average received power, ${\bf{H}}$ is the ${N_{\rm{r}}} \times {N_{\rm{t}}}$ narrow-band millimeter wave channel matrix, ${{\textbf{F}}_{{\rm{RF}}}}$ of size $N_{\rm{t}}\times N_{\rm{RF}}^{\rm{t}}$ is the analog TPC matrix, ${{\textbf{F}}_{{\rm{BB}}}}$ of size $N_{\rm{RF}}^{\rm{t}}\times N_s$ is the baseband TPC matrix, \textbf{s} is the $N_{\rm{s}}\times 1$ signal vector associated with $\mathbb{E}[{\bf{s}}{{\bf{s}}^*}] = \frac{1}{{{N_{\rm{s}}}}}{{\bf{I}}_{{N_{\rm{s}}}}}$, and ${\bf{n}}$ is the additive white Gaussian noise vector of independent and identically distributed (i.i.d.) ${\cal {CN}}(0,\sigma _{\rm{n}}^{2})$.

With the limited spatial scattering of millimeter wave signals, typically the geometric Saleh-Valenzuela model is used for modelling the millimeter wave channel, which is given by
\begin{equation}\label{2.4}
{\bf{H}} = \sqrt {\dfrac{{{N_{\rm{t}}}{N_{\rm{r}}}}}{{N_{\rm{cl}}{N_{\rm{ray}}}}}} \sum\limits_{m=1}^{N_{\rm{cl}}} {\sum\limits_{n=1}^{{N_{\rm{ray}}}}} {{\alpha _{m,n}}}  {{\bf{a}}_{\rm{r}}}(\theta _{m,n}^{\rm r}){{\bf{a}}_{\rm{t}}^*}(\theta _{m,n}^{\rm t}),
\end{equation}
where ${N_{\rm{cl}}}$ is the number of scattering clusters and each cluster contributes ${N_{\rm{ray}}}$ propagation paths, ${\alpha _{m,n}}$ denotes the complex gain of the $n^{th}$ path in the $m^{th}$ cluster, while $\theta _{m,n}^{\rm r}$ and $\theta _{m,n}^{\rm t} \in (0,2\pi ]$ are the AOA and AOD, respectively. ${{\bf{a}}_{\rm{t}}}(\theta _{m,n}^{\rm t})$ and ${{\bf{a}}_{\rm{r}}}(\theta _{m,n}^{\rm{r}})$ are the array response vectors of the transmitter and receiver, respectively. For an N-element uniform linear array (ULA), the array response vector can be written as
\begin{equation}\label{2.5}
\begin{array}{l}
{{\bf{a}}_{\rm{ULA}}}(\theta) = \dfrac{1}{{\sqrt {{N}} }}\Big[1,{\kern 1pt} {e^{j(2\pi /\lambda )d{\rm sin}(\theta)}},...,{e^{j({N} - 1)(2\pi /\lambda )d{\rm sin}(\theta)}}{\Big]^T},
\end{array}
\end{equation}
where $\lambda$ is the signal wavelength and $d = \lambda /2$ denotes the aperture domain sample spacing. %Note that the GSAC architecture considered and proposed schemes are also applicable to other types of antenna arrays, such as the uniform planar array (UPA) and the uniform circular array (UCA).

\section{Energy efficient Hybrid TPC}
In this section, we first describe the structure of the proposed GSAC architecture. Then, we propose a SIC-based hybrid TPC algorithm for maximizing the total achievable rate for arbitrary RF and antenna configurations. Finally, we analyse the energy efficiency of the FC, SAC and GSAC architectures, and propose an exhaustive search scheme for determining the RF and antenna configurations of our GSAC architecture having the highest energy efficiency. Note that the perfect CSI is tentatively assumed to be known at both the transmitter and receiver in this section. The situation when only the receiver knows the perfect CSI will be considered in Section IV.

\subsection{The GSAC architecture}
\begin{figure}[t]
  \centering
  \includegraphics[scale=0.48]{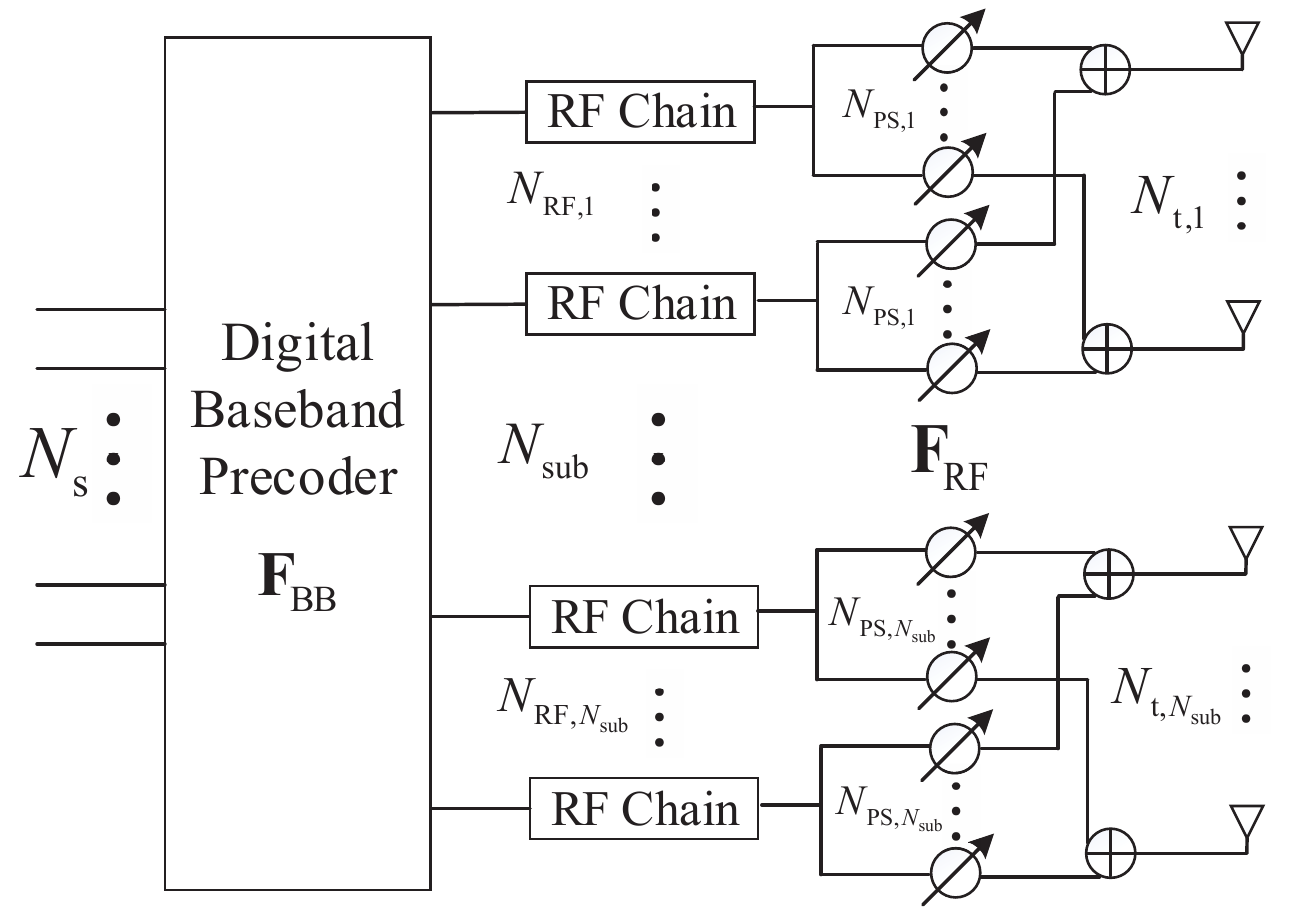}%fig3-Extended_Subarray.eps
  \caption{The GSAC architecture for the hybrid TPC.}
  \label{Extended_Subarray}
\end{figure}
Fig. \ref{Extended_Subarray} shows the transmitter of the considered GSAC architecture and the receiver has the `inverse architecture'. In the GSAC architecture, the number of RF chains connected to a sub-array and the number of antennas in each sub-array can be arbitrary. At the transmitter, the number of sub-arrays is denoted by $N_{\rm sub}$, while $N_{{\rm RF}, i}$  and $N_{{\rm t}, i}$ $, i = 1, 2, ..., N_{\rm sub}$ denote the number of RF chains and antennas connected to the $i^{th}$ sub-array, respectively. Furthermore, $N_{{\rm PS}, i}$ denotes the number of phase shifters connected to a single RF chain in the $i^{th}$ sub-array and ${N_{{\rm{PS}}}}$ is the total number of phase shifters. For the above parameters, we have the following relationships:
\begin{equation}\label{2.}
{N_{{\rm{RF}}}^{\rm t}}=\sum\limits_{i = 1}^{N_{\rm{sub}}} N_{{\rm{RF}},i},
\end{equation}
\begin{equation}\label{2.}
{N_{{\rm{t}}}}=\sum\limits_{i = 1}^{N_{\rm{sub}}} N_{{\rm{t}},i},
\end{equation}
\begin{equation}\label{2.}
{N_{{\rm{PS}}}}=\sum\limits_{i = 1}^{N_{\rm{sub}}}{N_{{\rm{PS}},i}}{N_{{\rm{RF}},i}}.
\end{equation}

Moreover, the following inequalities should also be satisfied,
\begin{equation}\label{2.}
1 \leq N_{{\rm{RF}},i} \leq N_{{\rm{t}},i} = {N_{{\rm{PS}},i}},
\end{equation}
\begin{equation}\label{2.}
{N_{\rm{sub}}} \leq {N_{{\rm{RF}}}^{\rm t}} \leq {N_{{\rm{t}}}} \leq {N_{{\rm{PS}}}},
\end{equation}
where (8) is a plausible conclusion, and (9) leads to
\begin{equation}\label{2.}
{N_{{\rm{RF}}}^{\rm t}}=\sum\limits_{i = 1}^{N_{\rm{sub}}} N_{{\rm{RF}},i} \geq \sum\limits_{i = 1}^{N_{\rm{sub}}} 1 = {N_{\rm{sub}}},
\end{equation}
\begin{equation}\label{2.}
{N_{{\rm{t}}}}=\sum\limits_{i = 1}^{N_{\rm{sub}}} N_{{\rm{t}},i} \geq \sum\limits_{i = 1}^{N_{\rm{sub}}} N_{{\rm{RF}},i} = {N_{{\rm{RF}}}^{\rm t}},
\end{equation}
\begin{equation}\label{2.}
{N_{{\rm{PS}}}}=\sum\limits_{i = 1}^{N_{\rm{sub}}}{N_{{\rm{PS}},i}}{N_{{\rm{RF}},i}} = \sum\limits_{i = 1}^{N_{\rm{sub}}}{N_{{\rm{t}},i}}{N_{{\rm{RF}},i}} \geq \sum\limits_{i = 1}^{N_{\rm{sub}}} N_{{\rm{t}},i} = {N_{{\rm{t}}}}.
\end{equation}

For any given RF and antenna configuration, the analog TPC matrix ${{\textbf{F}}_{{\rm{RF}}}}$ is a block-diagonal matrix, which can be expressed as
\begin{equation}\label{2.13}
{{\textbf{F}}_{{\rm{RF}}}} = \left[ \begin{array}{l}
{ {{\bf{F}}_{{{\rm{RF}}, {1}}}}}\\
{\kern 19pt}  \ddots  \\
{\kern 32pt} {{\bf{F}}_{{{\rm{RF}}, {N_{\rm{sub}}}}}}
\end{array} \right],
\end{equation}
where ${\bf{F}}_{{\rm{RF}}, {i}} = [{\bf{a}}_{{i, 1}}, {\bf{a}}_{{i, 2}}, ..., {\bf{a}}_{{i, N_{{\rm{RF}},i}}}],$
and ${\bf{a}}_{{i, j}}$  of size $N_{{\rm{t}},i} \times 1$ is the analog TPC vector for the $j^{th}$ RF chain in the $i^{th}$ sub-array. All the non-zero elements of ${\bf{F}}_{{\rm{RF}}, {i}}$ should satisfy the constant amplitude constraint, i.e., $\left| {\bf{F}}_{{\rm{RF}}, {i}} (\cdot , \cdot) \right| = 1/\sqrt {N_{{\rm{t}},i}}$. Moreover, the digital TPC matrix ${{\textbf{F}}_{{\rm{BB}}}}$ is also assumed to be a block-diagonal matrix similar as \cite{Gao2016}, i.e.,
\begin{equation}\label{2.14}
{{\textbf{F}}_{{\rm{BB}}}} = \left[ \begin{array}{l}
{ {{\bf{F}}_{{{\rm{BB}}, {1}}}}}\\
{\kern 19pt}  \ddots  \\
{\kern 32pt} {{\bf{F}}_{{{\rm{BB}}, {N_{{\rm{sub}}}}}}}
\end{array} \right],
\end{equation}
where ${\bf{F}}_{{\rm{BB}}, {i}} = [{\bf{d}}_{{i, 1}}; {\bf{d}}_{{i, 2}}; ...; {\bf{d}}_{{i,  N_{{\rm{RF}},i}}}]$ is used for cancelling the interference and for allocating power to the data streams on the $i_{th}$ sub-array and ${\bf{d}}_{{i, j}}$ is a $1 \times N_{{\rm{RF}},i}$ digital TPC vector. \emph{Note that $N_{\rm s} = N_{{\rm{RF}}}$ with equal power allocation is assumed}. To simplify the expression, we denote the hybrid TPC matrix by $\bf F = {{\bf{F}}_{{\mathop{\rm RF}\nolimits} }}{{\bf{F}}_{{\mathop{\rm BB}\nolimits} }}$ in the rest of the paper. It may be observed that $\bf F$ is also of block-diagonal structure and should satisfy ${\left\| {\bf{F}} \right\|_F}\leq N_{\rm s}$ to meet the total transmit power constraint.

\subsection{SIC-based hybrid TPC for the GSAC architecture}
In this subsection, we present the SIC based hybrid TPC scheme designed for any arbitrary RF and antenna configuration suitable for the GSAC architecture. Please note that the ultimate aim of this paper is to maximize the energy efficiency of the hybrid TPC. We have to accomplish the following two steps to achieve our aim. Firstly, the structure of the hybrid TPC depends on the specific RF and antenna configurations. Section III.B describes the first step, which aims for designing the hybrid TPC matrix capable of maximizing the total achievable rate. Then, the energy efficiency exhibits nonlinear dependence on the number of data streams, RF chains and antennas. Section III.C constitutes the next step, which carries out an exhaustive search for evaluating all the RF as well as antenna configurations and finds the one having the highest energy efficiency.

The total achievable rate can be expressed as
\begin{equation}\label{3.}
R={\log _2}\Big (\Big | {{\bf{I}}_{{N_{\rm{r}}}}}{\bf{ + }}\dfrac{\rho }{{{N_{\rm{s}}}\sigma^2}}{\bf{H}}{{\bf{F}}}{\bf{F}}^*{{\bf{H}}^{\bf{*}}}\Big |\Big ),
\end{equation}
and the rate optimization problem can be written as
\begin{equation}\label{3.15}
\begin{split}
{\bf{F}}^{\rm opt} = &\mathop{{\rm{arg}}{\kern 1pt}{\rm{max}}}\limits_{{{\bf{F}}}}R,\\
&{\kern 3pt}{\rm{s.t.}}{\kern 7pt}  {{\bf{F}}_{{\rm{RF}}}} \in {{\cal F}_{{\rm{RF}}}},\\
&{\kern 22pt}\left\| {{{\bf{F}}_{{\rm{RF}}}}{{\bf{F}}_{{\rm{BB}}}}} \right\|_{_F}^2 \leq {N_s},
\end{split}
\end{equation}
%\begin{equation}\label{3.16}
%R = {\log _2}\Big (\Big | {{\bf{I}}_{{N_{\rm{r}}}}}{\bf{ + }}\dfrac{\rho }{{{N_{\rm{s}}}\sigma^2}}{\bf{H}}{{\bf{F}}}{\bf{F}}^*{{\bf{H}}^{\bf{*}}}\Big |\Big ).
%\end{equation}
where ${{\cal F}_{{\rm{RF}}}}$ is the set of the feasible RF precoders induced by the constant amplitude constraint. With the non-convex constraint imposed on the analog TPC matrix ${\bf{F}}_{{\rm RF}}$, it is an open challenge to obtain the globally optimal solution to  (\ref{3.15}). However, by exploiting that the hybrid TPC matrix $\bf F$ is of block-diagonal structure, which implies that the TPC of each sub-array is mutually independent, the total achievable rate $R$ may be decomposed into multiple sub-rates, each of which is only related to a single sub-array. Then, we can solve (\ref{3.15}) by maximizing each of the sub-rates.

Note that when $ {N_{\rm{sub}}} \!=\! 1$, this is actually the fully-connected architecture, therefore we only consider cases when $ {N_{\rm{sub}}} \!>\! 1$.
The hybrid TPC matrix $\bf F $ may be partitioned as
\begin{numcases}{\!\!\!\!\!\!\!\!\!\bf F =}
\left[{\bf \hat f}_{{N_{\rm{sub}}}-1} \; {\bf f}_{{N_{\rm{sub}}}}\right], &\!\!\!\!\!\!\! if ${N_{{\rm{RF}}}^{\rm t}} = {N_{\rm{sub}}} = 2$, \\
\left[{\bf \hat F}_{{N_{\rm{sub}}}-1} \; {\bf f}_{{N_{\rm{sub}}}}\right], &\!\!\!\!\!\!\! if $N_{{\rm{RF}}\!,{N_{\rm{sub}}}} \!= \!1, {N_{{\rm{RF}}}^{\rm t}} \!>\! 2$, \\
\left[{\bf \hat f}_{{N_{\rm{sub}}}-1} \; {\bf F}_{{N_{\rm{sub}}}}\right], &\!\!\!\!\!\!\! if $N_{{\rm{RF}}\!,{N_{\rm{sub}}}} \!=\! {N_{{\rm{RF}}}^{\rm t}}\!\!-\!\!1 \!>\! 1$, \\
\left[{\bf \hat F}_{{N_{\rm{sub}}}-1} \; {\bf F}_{{N_{\rm{sub}}}}\right], &\!\!\!\!\!\!\! others,
\end{numcases}
where ${\bf f}_{{N_{\rm{sub}}}}$ \!\!or\! ${\bf F}_{{N_{\rm{sub}}}}$\!\! are the last $N_{{\rm{RF}}\!,{N_{\rm{sub}}}}$ \!\!columns of $\bf F $, and ${\bf \hat f}_{{N_{\rm{sub}}}\!-1}$ \!or\! ${\bf \hat F}_{{N_{\rm{sub}}}\!-1}$ \!\!represent the first ${N_{\rm{sub}}}\!\!\!-\!\!1$\! hybrid TPCs. We then consider the case presented in (20) as an example to introduce the rate decomposition processes, while the corresponding operations for other cases can be carried out similarly. The total achievable rate $R$  can be rewritten as
\begin{equation}\label{3.21}
\small
\begin{split}
R &={\log _2}\Big (\Big | {{\bf{I}}_{{N_{\rm{r}}}}}{\bf{ + }}\dfrac{\rho }{{{N_{\rm{s}}}\sigma^2}}{\bf{H}}{{\bf{F}}}{\bf{F}}^*{{\bf{H}}^{\bf{*}}}\Big |\Big )\\
&={\log _2}\Big (\Big | {{\bf{I}}_{{N_{\rm{r}}}}}{\bf{ + }}\dfrac{\rho }{{{N_{\rm{s}}}\sigma^2}}{\bf{H}}{\left[{\bf \hat F}_{{N_{\rm{sub}}}-1} \; {\bf F}_{{N_{\rm{sub}}}}\right]}{\left[{\bf \hat F}_{{N_{\rm{sub}}}-1} \; {\bf F}_{{N_{\rm{sub}}}}\right]}^*{{\bf{H}}^{\bf{*}}}\Big |\Big )\\
&={\log _2}\Big (\Big | {{\bf{I}}_{{N_{\rm{r}}}}}{\bf{ + }}\dfrac{\rho }{{{N_{\rm{s}}}\sigma^2}}{\bf{H}}{\bf \hat F}_{{N_{\rm{sub}}}-1}{\bf \hat F}_{{N_{\rm{sub}}}-1}^*{{\bf{H}}^{\bf{*}}}\\
&{\kern 35pt} +\dfrac{\rho }{{{N_{\rm{s}}}\sigma^2}}{\bf{H}}{\bf F}_{{N_{\rm{sub}}}}{\bf F}_{{N_{\rm{sub}}}}^*{{\bf{H}}^{\bf{*}}}\Big |\Big )\\
&\overset{(a)}{=}{\log _2}(|{{\bf C}_{{N_{\rm{sub}}}-1}} |)\\
&{\kern 35pt} + {\log _2}\Big (\Big |{{\bf{I}}_{{N_{\rm{r}}}}} + \dfrac{\rho }{{{N_{\rm{s}}}\sigma^2}}{{\bf C}^{-1}_{{N_{\rm{sub}}}-1}}{\bf{H}}{\bf F}_{{N_{\rm{sub}}}}{\bf F}_{{N_{\rm{sub}}}}^*{{\bf{H}}^{\bf{*}}}\Big |\Big )\\
&\overset{(b)}{=}{\log _2}(|{{\bf C}_{{N_{\rm{sub}}}-1}} |)\\
&{\kern 20pt} + {\log _2}\Big (\Big |{{\bf{I}}_{N_{{\rm{RF}},{N_{\rm{sub}}}}}} + \dfrac{\rho }{{{N_{\rm{s}}}\sigma^2}}{\bf F}^*_{{N_{\rm{sub}}}}{\bf{H}}^*{{\bf C}^{-1}_{{N_{\rm{sub}}}-1}}{{\bf{H}}}{\bf F}_{{N_{\rm{sub}}}}\Big |\Big )\\
&={\log _2}(|{{\bf C}_{{N_{\rm{sub}}}-2}} |)\\
&{\kern 20pt} + {\log _2}\Big (\Big |{{\bf{I}}_{N_{{\rm{RF}},{N_{\rm{sub}}}}}} + \dfrac{\rho }{{{N_{\rm{s}}}\sigma^2}}{\bf F}^*_{{N_{\rm{sub}}}}{\bf{H}}^*{{\bf C}^{-1}_{{N_{\rm{sub}}}-1}}{{\bf{H}}}{\bf F}_{{N_{\rm{sub}}}}\Big |\Big )\\
&{\kern 2pt} + {\log _2}\Big (\Big |{{\bf{I}}_{N_{{\rm{RF}},{N_{\rm{sub}}}-1}}} + \dfrac{\rho }{{{N_{\rm{s}}}\sigma^2}}{\bf F}^*_{{N_{\rm{sub}}}-1}{\bf{H}}^*{{\bf C}^{-1}_{{N_{\rm{sub}}}-2}}{{\bf{H}}}{\bf F}_{{N_{\rm{sub}}}-1}\Big |\Big )\\
&{\kern 5pt}\vdots\\
&\overset{(c)}{=}\sum\limits_{i = 1}^{N_{\rm{sub}}}{\log _2}\Big (\Big |{{\bf{I}}_{N_{{\rm{RF}},i}}} + \dfrac{\rho }{{{N_{\rm{s}}}\sigma^2}}{\bf F}^*_{i}{\bf{H}}^*{{\bf C}^{-1}_{i-1}}{{\bf{H}}}{\bf F}_{i}\Big |\Big ),
\end{split}
\end{equation}
where ${{\bf C}_{i-1}} = {{\bf{I}}_{{N_{\rm{r}}}}}{\bf{ + }}\frac{\rho }{{{N_{\rm{s}}}\sigma^2}}{\bf{H}}{\bf \hat F}_{i-1}{\bf \hat F}_{i-1}^*{{\bf{H}}^{\bf{*}}}$, ${\bf C}_0 = {\bf I}_{N_{{\rm{RF}},1}}$. Step (a) is true due to the fact that we have $|\bf AB | = |\bf A||\bf B|$ and we let ${\bf A} = {\log _2}(|{{\bf C}_{{N_{\rm{sub}}}-1}} |)$, ${\bf B} = {\log _2}\Big (\Big |{{\bf{I}}_{{N_{\rm{r}}}}} + \frac{\rho }{{{N_{\rm{s}}}\sigma^2}}{{\bf C}^{-1}_{{N_{\rm{sub}}}-1}}{\bf{H}}{\bf F}_{{N_{\rm{sub}}}}{\bf F}_{{N_{\rm{sub}}}}^*{{\bf{H}}^{\bf{*}}}\Big |\Big )$. Furthermore, (b) is obtained due to the fact that $|\bf I + AB | = |I + BA|$ by defining ${\bf A} = {{\bf C}^{-1}_{{N_{\rm{sub}}}-1}}{\bf{H}}{\bf F}_{{N_{\rm{sub}}}}$ and ${\bf B} = {\bf F}_{{N_{\rm{sub}}}}^*{{\bf{H}}^{\bf{*}}}$. Note that the second term ${\log _2}\Big (\Big |{{\bf{I}}_{N_{{\rm{RF}},{N_{\rm{sub}}}}}} + \frac{\rho }{{{N_{\rm{s}}}\sigma^2}}{\bf F}^*_{{N_{\rm{sub}}}}{\bf{H}}^*{{\bf C}^{-1}_{{N_{\rm{sub}}}-1}}{{\bf{H}}}{\bf F}_{{N_{\rm{sub}}}}\Big |\Big )$ of step (b) is the achievable sub-rate of the $({{N_{\rm{sub}}}}){th}$ sub-array and the form of the first term ${\log _2}(|{{\bf C}_{{N_{\rm{sub}}}-1}} |)$ is similar to $R$. This observation implies that we can further decompose ${\log _2}(|{{\bf C}_{{N_{\rm{sub}}}-1}} |)$ utilizing a similar method to that in (\ref{3.21}). Step (c) represents the result after ${{N_{\rm{sub}}}}$ decompositions.

As has shown in (\ref{3.21}), the total achievable rate $R$ is the sum of the sub-rates of all the sub-arrays. Therefore, the total rate-optimization problem of (\ref{3.15}) can be transformed into a series of sub-rate optimization problems for the sub-arrays, which can be solved one by one. Similar to \cite{Gao2016}, we adopt the idea of SIC to optimize all the sub-rates. The sub-rate optimization problem of the $i^{th}$ sub-array can be formulated as
\begin{equation}\label{3.22}
{\bf F}^{\rm opt}_{i} = \mathop {{\rm{arg}}{\kern 1pt}{\rm{max}}}\limits_{{\bf F}_{i}\in {{\cal F}_{{{i}}}}}{\log _2}\Big (\Big |{{\bf{I}}_{N_{{\rm{RF}},i}}} + \dfrac{\rho }{{{N_{\rm{s}}}\sigma^2}}{\bf F}^*_{i}{{\bf P}_{i-1}}{\bf F}_{i}\Big |\Big ),
\end{equation}
where ${{\cal F}_{{{i}}}}$ includes all feasible vectors satisfying the transmit power constraint and the constant amplitude constraint on the analog precoder, and ${{\bf P}_{i-1}} = {\bf{H}}^*{{\bf C}^{-1}_{i-1}}{{\bf{H}}}$ is an ($N_{\rm t} \times N_{\rm t}$) Hermitian matrix. Note that only the elements spanning from the $A^{th} = (\sum\limits_{j = 1}^{{{i-1}}} N_{{\rm{t}},j}+ 1)^{th}$ row to the $B^{th} = (\sum\limits_{j = 1}^{{{i}}} N_{{\rm{t}},j})^{th}$ row within ${\bf F}_{i} $ are non-zero (${N_{{\rm{t}},0}}$ is set to be 0). Therefore, the sub-rate optimization problem (\ref{3.22}) can be written as
\begin{equation}\label{3.23}
{\bf\widetilde F}^{\rm opt}_{i} = \mathop {{\rm{arg}}{\kern 1pt}{\rm{max}}}\limits_{{\bf\widetilde F}_{i}\in {{\cal \widetilde F}_{{{i}}}}}{\log _2}\Big (\Big |{{\bf{I}}_{N_{{\rm{RF}},i}}} + \dfrac{\rho }{{{N_{\rm{s}}}\sigma^2}}{\bf \widetilde F}^*_{i}{{\bf\widetilde P}_{i-1}}{\bf\widetilde F}_{i}\Big |\Big ),
\end{equation}
where ${{\cal \widetilde F}_{{{i}}}}$ is the set of all possible ${N_{{\rm{t}},i}}\times 1$ vectors satisfying the transmit power constraint and the constant amplitude constraint on the analog precoder, ${\bf\widetilde F}_i$ of size ${N_{{\rm{t}},i}} \times {N_{{\rm{RF}},i}}$  is the sub-matrix of ${\bf F}_{i} $ spanning from the $A^{th}$ row to the $B^{th}$ row, ${\bf\widetilde P}_{i-1}$ of size ${N_{{\rm{t}},i}} \times {N_{{\rm{t}},i}}$  is the sub-matrix of ${\bf P}_{i-1}$ from the $A^{th}$ row and column to the $B^{th}$ row and column. Let us define the singular value decomposition (SVD) of the Hermitian matrix ${\bf\widetilde P}_{i-1}$ as
\begin{equation}\label{3.}
{\bf\widetilde P}_{i-1} = {\bf{V}}_{i-1}{\bf{\Sigma}}_{i-1}{\bf{V}^*_{i-1}},
\end{equation}
where ${\bf{\Sigma}}_{i-1}$ is a diagonal matrix containing the singular values of ${\bf\widetilde P}_{i-1}$ in decreasing order and ${\bf{V}}_{i-1}$ is a unitary matrix of size ${N_{{\rm{t}},i}} \times {N_{{\rm{t}},i}}$. It is widely exploited that the optimal unconstrained TPC matrix of the $i^{th}$ sub-array is constituted by the first ${N_{{\rm{RF}},i}}$ columns of ${\bf{V}}_{i-1}$, i.e.,
\begin{equation}\label{3.}
{\bf\widetilde F}^{\rm opt}_{i} = {\bf{V}}_{i-1}(: , 1 \!:\! {N_{{\rm{RF}},i}}).
\end{equation}
The total optimal unconstrained TPC matrix ${\bf F}^{\rm opt}$ is the block diagonal concatenation of ${\bf\hat F}^{\rm opt}_{i} $, which can be obtained through $N_{\rm sub} $ iterations formulated as
\begin{equation}\label{3.26}
{\bf F}^{\rm opt} = \left[ \begin{array}{l}
{\bf\widetilde F}^{\rm opt}_{1}\\
{\kern 19pt}  \ddots  \\
{\kern 32pt} {\bf\widetilde F}^{\rm opt}_{N_{\rm sub}}
\end{array} \right].
\end{equation}

Since there are constant amplitude constrains placed on the elements of the analog TPC matrix ${\bf F}_{\rm RF}$, we cannot directly set ${\bf F}^{\rm opt} $ as the solution of the optimization problem (\ref{3.15}). To obtain a practical solution, we try to further convert (\ref{3.23}).
\begin{lemma}
When ${N_{{\rm{RF}},i}} = 1$, the optimization problem (\ref{3.23}) can be rewritten as
\begin{equation}\label{3.}
{\bf\widetilde f}^{\rm opt}_{i} = \mathop {{\rm{arg}}{\kern 1pt}{\rm{max}}}\limits_{{\bf\widetilde f}_{i}\in {{\cal \widetilde F}_{{{i}}}}}{\log _2}\Big (\Big |{{\bf{I}}_{N_{{\rm{RF}},i}}} + \dfrac{\rho }{{{N_{\rm{s}}}\sigma^2}}{\bf \widetilde f}^*_{i}{{\bf\widetilde P}_{i-1}}{\bf\widetilde f }_{i}\Big |\Big ),
\end{equation}
which is approximately equivalent to
\begin{equation}\label{3.28}
{\bf\widetilde f}^{\rm opt}_{i} = \mathop {{\rm{arg}}{\kern 1pt}{\rm{min}}}\limits_{{\bf\widetilde f}_{i}\in {{\cal \widetilde F}_{{{i}}}}}\left\|{\bf v}_{i-1} - {\bf\widetilde f}_{i} \right\|^2_2,
\end{equation}
where ${\bf v}_{i-1}$ is the first right singular vector of ${\bf\hat P}_{i-1}$.
\end{lemma}
\begin{IEEEproof}
See Appendix A in \cite{Gao2016}.
\end{IEEEproof}
The solution to (\ref{3.28}) can be readily expressed as
\begin{equation}\label{3.}
{{\bf a}_{i}} = \frac {1}{\sqrt{N_{{\rm{t}},i}}}{\rm exp}[j{\rm angle}({\bf v}_{i-1})],\\
\end{equation}
\begin{equation}\label{3.}
{{\bf d}_{i}} = {\left\| {\bf v}_{i-1} \right\|_1}\big/{\sqrt{N_{{\rm{t}},i}}},
\end{equation}
\begin{equation}\label{3.31}
{{\bf\widetilde f}_{i}} = \frac {1}{{N_{{\rm{t}},i}}}{\left\| {\bf v}_{i-1} \right\|_1}{\rm exp}[j{\rm angle}({\bf v}_{i-1})],
\end{equation}
where ${\rm angle}({\bf v}_{i-1})$ denotes the phase vector of ${\bf v}_{i-1}$.
\begin{lemma}
When ${N_{{\rm{RF}},i}} > 1$, the optimization problem (\ref{3.23}) is approximately equivalent to
\begin{equation}\label{3.32}
{\bf\widetilde F}^{\rm opt}_{i} = \mathop {{\rm{arg}}{\kern 1pt}{\rm{min}}}\limits_{{\bf\widetilde F}_{i}\in {{\cal \widetilde F}_{{{i}}}}}\left\|{\bf{V}}_{i-1}(: , 1 \!:\! {N_{{\rm{RF}},i}})- {\bf\widetilde F}_{i} \right\|^2_F.
\end{equation}
\end{lemma}
\begin{IEEEproof}
The proof is similar to that of Section III in \cite{Ayach2014} and thus it is omitted.
\end{IEEEproof}

Similar to the solution of (\ref{3.28}), the practical analog/digital TPC matrices of (\ref{3.32}) are given by
\begin{equation}\label{3.}
{\bf{F}}_{{\rm{RF}}, {i}} = \frac {1}{\sqrt{N_{{\rm{t}},i}}}{\rm exp}[j{\rm angle}({\bf{V}}_{i-1}(: , 1 \!:\! {N_{{\rm{RF}},i}}))],
\end{equation}
\begin{equation}\label{3.}
{\bf{F}}_{{\rm{BB}}, {i}} = ({\bf{F}}^*_{{\rm{RF}}, {i}}{\bf{F}}_{{\rm{RF}}, {i}})^{-1}{\bf{F}}^*_{{\rm{RF}}, {i}}{\bf{V}}_{i-1}(: , 1 \!:\! {N_{{\rm{RF}},i}}),
\end{equation}
\begin{equation}\label{3.35}
{{\bf\widetilde{F}}_i} =  {\bf{F}}_{{\rm{RF}}, {i}} {\bf{F}}_{{\rm{BB}}, {i}},
\end{equation}
where ${\bf{F}}_{{\rm{RF}}, {i}}$ extracts phases of elements in ${\bf{V}}_{i-1}(: , 1 \!:\! {N_{{\rm{RF}},i}})$ to satisfy the constant amplitude constraint and ${\bf{F}}_{{\rm{BB}}, {i}}$ is calculated by least squares. It is readily seen that both (\ref{3.31}) and (\ref{3.35}) satisfy the total transmit power constraint.

After we solve all the sub-rate optimization problems for all sub-arrays having either ${N_{{\rm{RF}},1}} = 1$ or ${N_{{\rm{RF}},1}} > 1$ RF chains, the total practical hybrid TPC matrix can be expressed as
\begin{equation}\label{3.36}
\small
{\bf{F}} = \left[ \begin{array}{l}
{ {{\bf\widetilde{F}}_{{{1}}}}}, \:{\rm if} \: {N_{{\rm{RF}},1}} > 1\\
{\kern 8pt}  \ddots  \\
{\kern 20pt} { {{\bf\widetilde{f}}_{{ {i}}}}}, \:{\rm if} \: {N_{{\rm{RF}},i}} = 1\\
{\kern 28pt}  \ddots  \\
{\kern 45pt} { {{\bf\widetilde{F}}_{{ {j}}}}}, \:{\rm if} \: {N_{{\rm{RF}},j}} > 1\\
{\kern 53pt}  \ddots  \\
{\kern 70pt} {{\bf\widetilde{f}}_{{{N_{\rm{sub}}}}}}, \:{\rm if} \: {N_{{\rm{RF}},{N_{\rm{sub}}}}} = 1
\end{array} \right].
\end{equation}

%As described above, the total achievable rate is decomposed into a series sub-rates and the total rate-optimization problem of (\ref{3.15}) is correspondingly decomposed into multiple sub-rate optimization problems, each of which is only related to a single subarray.
All the above details are summarized in {\bf Algorithm 1}.

\begin{algorithm}[t]
    \renewcommand{\algorithmicrequire}{\textbf{Input:}}
	\renewcommand{\algorithmicensure}{\textbf{Output:}}
   \renewcommand{\algorithmicrequire}{\textbf{Initialization:}}
	\caption{SIC Based Hybrid TPC for the GSAC Architecture}
    \label{alg:1}
	\begin{algorithmic}[1]
        \REQUIRE %input
        ${\bf H}$, ${N_{{\rm{RF}}}^{\rm t}}$, ${N_{{\rm{RF}},i}}$, ${N_{{\rm{t}},i}}, i = 1, 2, ..., {N_{\rm{sub}}}$
        \ENSURE The total hybrid TPC $\bf F$
        \STATE ${\bf P}= {\bf{H}}^*{{\bf{H}}}$
        \FOR {$i\leq {N_{\rm{sub}}}$}
        \STATE ${\bf\widetilde P}= {\bf{V}}{\bf{\Sigma}}{\bf{V}^*}$
        \IF {${N_{{\rm{RF}},i}} == 1$}
        \STATE  ${{\bf a}_{i}} = \frac {1}{\sqrt{N_{{\rm{t}},i}}}{\rm exp}[j{\rm angle}({\bf v}_{i-1})]$, ${{\bf d}_{i}} = \frac {\left\| {\bf v}_{i-1} \right\|_1}{\sqrt{N_{{\rm{t}},i}}}$
        \STATE ${\bf\widetilde f}_{i} = \frac {1}{{N_{{\rm{t}},i}}}{\left\| {\bf v}_{i-1} \right\|_1}{\rm exp}[j{\rm angle}({\bf v}_{i-1})]$
        \ELSE
        \STATE ${\bf{F}}_{{\rm{RF}}, {i}} = \frac {1}{\sqrt{N_{{\rm{t}},i}}}{\rm exp}[j{\rm angle}({\bf{V}}(: , 1 \!:\! {N_{{\rm{RF}},i}}))]$
        \STATE ${\bf{F}}_{{\rm{BB}}, {i}} = ({\bf{F}}^*_{{\rm{RF}}, {i}}{\bf{F}}_{{\rm{RF}}, {i}})^{-1}{\bf{F}}^*_{{\rm{RF}}, {i}}{\bf{V}}(: , 1 \!:\! {N_{{\rm{RF}},i}})$
        \STATE  ${\bf\widetilde F}_{i} = {\bf{F}}_{{\rm{RF}}, {i}} {\bf{F}}_{{\rm{BB}}, {i}}$
        \ENDIF
        \IF {$i == 1$ and ${N_{{\rm{RF}},1}} == 1$}
        \STATE  ${{\bf C}_{i}} = {{\bf{I}}_{{N_{\rm{r}}}}}{\bf{ + }}\frac{\rho }{{{N_{\rm{s}}}\sigma^2}}{\bf{H}}{\bf \hat f}_{i}{\bf \hat f}_{i}^*{{\bf{H}}^{\bf{*}}}$
        \ELSE
        \STATE ${{\bf C}_{i}} = {{\bf{I}}_{{N_{\rm{r}}}}}{\bf{ + }}\frac{\rho }{{{N_{\rm{s}}}\sigma^2}}{\bf{H}}{\bf \hat F}_{i}{\bf \hat F}_{i}^*{{\bf{H}}^{\bf{*}}}$
        \ENDIF
        \STATE Update ${\bf P} = {\bf{H}}^*{{\bf C}^{-1}_{i}}{{\bf{H}}}$
        \ENDFOR
        \STATE Construct $\bf F$ as (\ref{3.36})
    \end{algorithmic}
\end{algorithm}

\subsection{Energy efficiency}
The energy efficiency can be defined as the ratio of the achievable rate and of the total power consumption \cite{Yu2016}, i.e.,
\begin{equation}\label{4.4}
\eta = \frac{R}{P_{\rm total}} = \frac{R}{{{P_{\rm CO}} + {N_{{\rm{RF}}}^{\rm t}}{P_{{\rm{RF}}}} + {N_{\rm{t}}}{P_{{\rm{PA}}}} + {N_{{\rm{PS}}}}{P_{{\rm{PS}}}}}},
\end{equation}
where ${P_{\rm CO}}$ is the common power of the transmitter including site-cooling, baseband processing and synchronization. ${P_{{\rm{RF}}}}$, ${P_{{\rm{PA}}}} $, and ${P_{{\rm{PS}}}}$ represent the power consumption of each RF chain, power amplifier and phase shifter, respectively.
Note that ${N_{\rm{t}}}$ and $N_{{\rm{RF}}}$ are constants, only ${N_{{\rm{PS}}}}$ is a variable.
The number of phase shifters for the FC, SAC, and GSAC architectures are summarized as follows.
\begin{numcases}{\!\!\!\!\!\!\!\!\!N_{{\rm{PS}}} =}
{N_{\rm{t}}}{N_{{\rm{RF}}}^{\rm t}}, & \!\!\!\!\!\!\!FC , \\
{N_{\rm{t}}}, & \!\!\!\!\!\!\!SAC , \\
\sum\limits_{i = 1}^{N_{\rm{sub}}}{N_{{\rm{PS}},i}}{N_{{\rm{RF}},i}}, & \!\!\!\!\!\!\!\!GSAC,
\end{numcases}
which satisfy the following inequality
\begin{equation}\label{4.}
{N_{\rm{t}}} \leq \sum\limits_{i = 1}^{N_{\rm{sub}}}{N_{{\rm{PS}},i}}{N_{{\rm{RF}},i}} \leq {N_{\rm{t}}}{N_{{\rm{RF}}}^{\rm t}},
\end{equation}
where the first inequality has been given in (12), while the second inequality is given by
\begin{equation}\label{4.}
\sum\limits_{i = 1}^{N_{\rm{sub}}}{N_{{\rm{PS}},i}}{N_{{\rm{RF}},i}} = \sum\limits_{i = 1}^{N_{\rm{sub}}}{N_{{\rm{t}},i}}{N_{{\rm{RF}},i}} \leq \sum\limits_{i = 1}^{N_{\rm{sub}}}{N_{{\rm{t}},i}}{N_{{\rm{RF}}}^{\rm t}} = {N_{\rm{t}}}{N_{{\rm{RF}}}^{\rm t}}.
\end{equation}
Therefore, we can conclude that $ {P^{\rm SAC}_{\rm total}} \leq {P^{\rm GSAC}_{\rm total}} \leq {P^{\rm FC}_{\rm total}}$.

Moreover, to simplify the analysis, we assume that the number of antennas connected to a sub-array is proportional to the number of RF chains in this sub-array, i.e.,
\begin{equation}\label{4.42}
N_{{\rm{t}},i} = \frac{N_{{\rm{t}}}}{N_{{\rm{RF}}}}N_{{\rm{RF}},i}.
\end{equation}
Then, we directly arrive at
\begin{equation}\label{4.43}
\frac{N_{{\rm{t}}}}{N_{{\rm{RF}}}^{\rm t}} \leq N_{{\rm{t}},i} \leq {N_{{\rm{t}}}}.
\end{equation}
Generally, the more antennas are connected to each RF chain, the higher the achievable rate becomes. Therefore, in terms of the achievable rate, we have $ {R^{\rm SAC}} \leq {R^{\rm GSAC}} \leq {R^{\rm FC}}$.

Based on the above analysis, we infer that the FC architecture has the highest achievable rate at the cost of the highest power consumption. By contrast, the SAC architecture has the lowest power consumption at the cost of a reduced rate. Due to the nonlinear dependence of the energy efficiency on the number of data streams, RF chains and antennas, the architecture having highest energy efficiency is theoretically unknown. To solve the above problem, we can afford using an exhaustive search scheme to evaluate all the RF and antenna configurations and select the one with best energy efficiency. The numerical comparisons of Section V demonstrate that our GSAC architecture is capable of achieving the best energy efficiency in the family of FC and SAC architectures.

\begin{algorithm}[t]
    \renewcommand{\algorithmicrequire}{\textbf{Input:}}
	\renewcommand{\algorithmicensure}{\textbf{Output:}}
   \renewcommand{\algorithmicrequire}{\textbf{Initialization:}}
	\caption{Exhaustive Search Scheme }
    \label{alg:2}
	\begin{algorithmic}[1]
        \REQUIRE %input
        ${N_{{\rm{RF}}}^{\rm t}}$
        \ENSURE ${N_{{\rm{RF}},i}}$, $i = 1, 2, ..., {N_{\rm{sub}}}$
        \STATE Generate all the possible RF configuration $\mathcal S $ for the given ${N_{{\rm{RF}}}}$ by generating function \cite{Abramowitz1965}.
        \FOR {Each RF configuration in $\mathcal S $}
        \STATE Calculate the achievable rate through \textbf{Algorithm \ref{alg:1}};
        \STATE Calculate the energy efficiency by (\ref{4.4});
        \ENDFOR
        \STATE Compare all the energy efficiencies and determine the configuration with best energy efficiency.
    \end{algorithmic}
\end{algorithm}

\emph{Remark}: Again, since the millimeter wave channel exhibits limited scattering and the total number of RF chains is small in reality, the total number of configurations is limited. Moreover, inspired by (\ref{3.21}), indicating that the total achievable rate $R$ is the sum of the sub-rates of all the sub-arrays, the configurations (1, 2, 5), (5, 1, 2) and (2, 1, 5) can be regarded as the same. The remaining problem is now how to calculate the total number and determine what all the possible configurations are. This is actually an \emph{integer partitioning issue}, i.e. writing $n$ as a sum of positive integers. Two sums that differ only in the order of their summands are considered to be the same partition. The integer partitioning can be readily solved by relying on the generating function or recursive technique of \cite{Abramowitz1965}, so as to obtain all the possible RF configurations. {The total number of possible partitions of a non-negative integer $n$ has an asymptotic expression of $p(n)\sim \frac{1}{4n\sqrt{3}}e^{\pi\sqrt{\frac{2n}{3}}}$ } \cite{Pak2006}. We summarize the total number of possible RF configurations $p(N_{\rm RF})$ in Table \ref{Table1}.
\begin{table}[!t]
\scriptsize
\renewcommand\arraystretch{2}
\centering
  \caption{THE TOTAL NUMBER OF POSSIBLE RF CONFIGURATIONS}
\begin{tabular}{|c|c|c|c|c|}\hline
  $N_{\rm RF}$ & 2 & 4 & 8 & 16 \\ \hline
  $p(N_{\rm RF})$ & 2 & 5 & 22 & 231 \\ \hline
\end{tabular}\label{Table1}
\end{table}
As we can see, even for $N_{\rm RF}=16$, $p(16)=231$ is not excessive. Therefore, the exhaustive search scheme is not computationally demanding.

For a given total number $N_{\rm RF}$ of RF chains, the set $\mathcal S $ contains all the possible RF configurations. For example, when $N_{\rm RF}^{\rm t} =4$, we have ${\mathcal S}_4 = \{(4), (3, 1), (2, 2), (2, 1, 1), (1, 1, 1, \\1) \}$. In the proposed exhaustive search scheme, we evaluate all the RF configurations in $\mathcal S $ one by one and finally adopt the one having the best energy efficiency. All the details are summarized in \textbf{Algorithm \ref{alg:2}}. {Please note that when the channel matrix $\bf{H}$ changes, the hybrid TPC has to be re-constructed. However, we observe from simulation results that the optimal RF configuration is robust.}

%#include<stdio.h>
%void main()
%{
%   int equation(int n,int m);
%   printf("quantity:%d\n",equation(32,32));
%}
%int equation(int n,int m)
%{
%   if(n==1||m==1)
%      return (1);
%   else if(n<m)
%      return equation(n,n);
%   else if(n==m)
%      return 1+equation(n,n-1);
%   else
%      return equation(n-m,m)+equation(n,m-1);
%}

\section{TPC to facilitate limited feedback}
\begin{algorithm}[t]
    \renewcommand{\algorithmicrequire}{\textbf{Input:}}
	\renewcommand{\algorithmicensure}{\textbf{Output:}}
   \renewcommand{\algorithmicrequire}{\textbf{Initialization:}}
	\caption{Beamsteering Codebook Based Hybrid TPC for the GSAC Architecture }
    \label{alg:3}
	\begin{algorithmic}[1]
        \REQUIRE %input
        $b$, ${N_{{\rm{RF}},i}}$, ${N_{{\rm{t}},i}}, i = 1, 2, ..., {N_{\rm{sub}}}$
        \ENSURE ${\bf F_{\rm RF}}$, ${\bf F_{\rm BB}}$
        \STATE Obtain the optimal unconstrained TPC matrix ${\bf F}^{\rm opt}$ as (\ref{3.26})
        \STATE Construct the beamsteering codebooks ${\bf A^{\rm quant}}$ as (\ref{4.46})
        \FOR {$i\leq {N_{\rm{sub}}}$}
        \FOR {$m\leq {N_{\rm{RF},i}}$}
        \STATE ${\bf F}_{{\rm RF},i}(:,m)=\mathop {\rm{max}}\limits_{m}\langle {\bf A}_i(:,n), {\bf\widetilde F}^{\rm opt}_{i}(:,m)\rangle$
        \ENDFOR
        \ENDFOR
        \STATE Set ${\bf F_{\rm RF}}$ as (\ref{2.13})
        \STATE Obtain ${\bf F}_{{\rm BB},i}$ using RVQ \cite{Alkhateeb2015} or MUBs \cite{Hanzo2018}
        \STATE Construct ${\bf F_{\rm BB}}$ as (\ref{2.14})
    \end{algorithmic}
\end{algorithm}
Section III considers the hybrid precoding for the proposed energy-efficient GSAC architecture when the perfect CSI is assumed at both the transmitter and the receiver. However, in practice especially in the millimeter wave MIMO system equipping with large antenna array, feeding back the perfect CSI form the receiver to the transmitter is difficult and will cause significant feedback overhead. Therefore, in this section, we consider the design of hybrid precoding scheme for the GSAC architecture to facilitate limited feedback, when the perfect CSI is only known at the receiver.

The basic idea is to design the quantization codebook for the GSAC architecture. When the receiver obtain the TPC, it quantizes TPC through the codebook and then feeds back to the transmitter. For the analog TPC, we adopt the beamsteering codebooks which are of relatively small size and make full use of angular domain information \cite{Alkhateeb2015}. Since the antenna structure of the GSAC architecture is different from the FC and SAC architectures, the constructed codebooks are also different.
Given an RF configuration, i.e., ${N_{{\rm{RF}},i}}, i = 1, 2, ..., {N_{\rm{sub}}}$, the antenna indexes are
\begin{equation}\label{4.5}
\begin{split}
{\bm\lambda}_{{\rm{t}}, 1}&=[{0,...,N_{\rm{t},1}-1}],\\
{\bm\lambda}_{{\rm{t}}, 2}&=\bigg[{N_{\rm{t},1},...,\sum\limits_{i = 1}^{2}{N_{{\rm{t}},i}}-1}\bigg],\\
\vdots \\
{\bm\lambda}_{{\rm{t}}, {N_{{\rm{sub}}}}}&=\Bigg[{\sum\limits_{i = 1}^{N_{\rm{sub}}-1}{N_{{\rm{t}},i}},...,\sum\limits_{i = 1}^{N_{\rm{sub}}}{N_{{\rm{t}},i}}-1}\Bigg],
\end{split}
\end{equation}
where $\Lambda_{{\rm{t}}i}$ denote the partitioned subset of antenna indexes connected to the $i^{th}$ sub-array. Therefore, the total antenna index matrix can be represented as
\begin{equation}\label{4.45}
\bf\Lambda_{{\rm{t}}}= \left[ \begin{array}{l}
{\bm\lambda}_{{\rm{t}}, 1}\\
{\kern 14pt}  \ddots  \\
{\kern 28pt} {\bm\lambda}_{{\rm{t}}, {N_{{\rm{sub}}}}}
\end{array} \right].
\end{equation}
Denoting the number of quantization bits by ${b}$, the candidate beamsteering matrix for the $i^{th}$ sub-array is ${\bf A}_i=\frac {1}{{N_{{\rm{t}},i}}}{\rm exp}[j\pi{\bm\lambda}^*_{{\rm{t}}, i}{\rm sin}({{\bf\theta}_b})]$, where ${{\bf \theta}_b}=[0, \frac{2\pi}{2^b}, ..., \frac{(2^b-1)2\pi}{2^b} ]$ is the candidate angle vector. Therefore, the quantized beamsteering codebooks for the GSAC architecture can be expressed as
\begin{equation}\label{4.46}
{\bf A^{\rm quant}}=\left[ \begin{array}{l}
{\bf A}_1\\
{\kern 14pt}  \ddots  \\
{\kern 28pt} {\bf A}_{N_{{\rm{sub}}}}
\end{array} \right].
\end{equation}
The receiver then selects the $N_{\rm RF}$ columns form ${\bf A^{\rm quant}}$ which exhibit maximum correlation with the columns of ${\bf F}^{\rm opt}$ and constructs the analog TPC ${\bf F}_{\rm RF}$ satisfying the constant amplitude constraint, i.e.,
\begin{equation}\label{4.47}
\begin{split}
&{\bf F}_{\rm RF}(:,m)=\mathop {\rm{max}}\limits_{m}\langle {\bf A^{\rm quant}}(:,n), {\bf F}^{\rm opt}(:,m)\rangle,\\
&{\kern 20pt}1\leq m \leq {N_{{\rm{RF}}}^{\rm t}}, 1\leq n \leq N_{\rm sub}2^b,
\end{split}
\end{equation}
Inspired by the block diagonal structure of both ${\bf A^{\rm quant}}$ and ${\bf F}^{\rm opt}$, the quantization procedures for the analog TPC of different sub-arrays can be separated and the problem (\ref{4.47}) can be simplified as
\begin{equation}\label{4.48}
\begin{split}
&{\bf F}_{{\rm RF},i}(:,m)=\mathop {\rm{max}}\limits_{m}\langle {\bf A}_i(:,n), {\bf\widetilde F}^{\rm opt}_{i}(:,m)\rangle,\\
&{\kern 5pt}1\leq i \leq N_{\rm sub}, 1\leq m \leq N_{{\rm RF},i}, 1\leq n \leq 2^b,
\end{split}
\end{equation}
As for the digital TPC, since its dimension is small and the corresponding quantization has been well studied, such as random vector quantization (RVQ) \cite{Alkhateeb2015} and mutually unbiased bases (MUBs) \cite{Hanzo2018}, we omit these procedures.

The pseudo code of the proposed beamsteering codebook based hybrid TPC scheme for the GSAC architecture is summarized in \textbf{Algorithm \ref{alg:3}}.

\section{Simulation Results}
In this section, the performance of the proposed SIC based hybrid TPC scheme (marked as GSAC-SIC) and the proposed beamsteering codebook based hybrid TPC scheme (marked as GSAC-codebook) for the considered GSAC architecture are evaluated. We adopt the orthogonal matching pursuit scheme in the FC architecture (marked as FC-OMP) \cite{Ayach2014} and the SIC based scheme in the SAC architecture (marked as SAC-SIC) \cite{Gao2016} as the benchmarks. The unconstrained TPC scheme for the GSAC architecture (marked as GSAC-opt) is given in (\ref{3.26}) and the scenario when the number of RF chains in the different sub-arrays is the same is labelled as ``GSAC-SIC-equal-RF". Both the transmitter and the receiver are equipped with ULAs with $\lambda/2 $ aperture domain sample spacing. Moreover, the channel parameters are set as $N_{\rm cl} = 10$ and $N_{\rm ray} = 5$. The azimuth AOAs and AODs obey the Laplacian distribution with uniformly distributed mean angles within $(0, 2\pi]$ and angular spread of $7.5^\circ$ \cite{Ayach2014}. The power consumptions of the different components are set as follows: $P_{\rm CO} = 10 {\rm W}$, $P_{\rm RF} = 100 {\rm mW}$, $P_{\rm PA} = 100 {\rm mW}$,  $P_{\rm PS} = 10 {\rm mW}$ \cite{Yu2016}. Finally, the signal-to-noise ratio (SNR) is defined as $\frac{\rho }{{\sigma^2}}$.

\begin{figure}[t]
  \centering
  \includegraphics[scale=0.5]{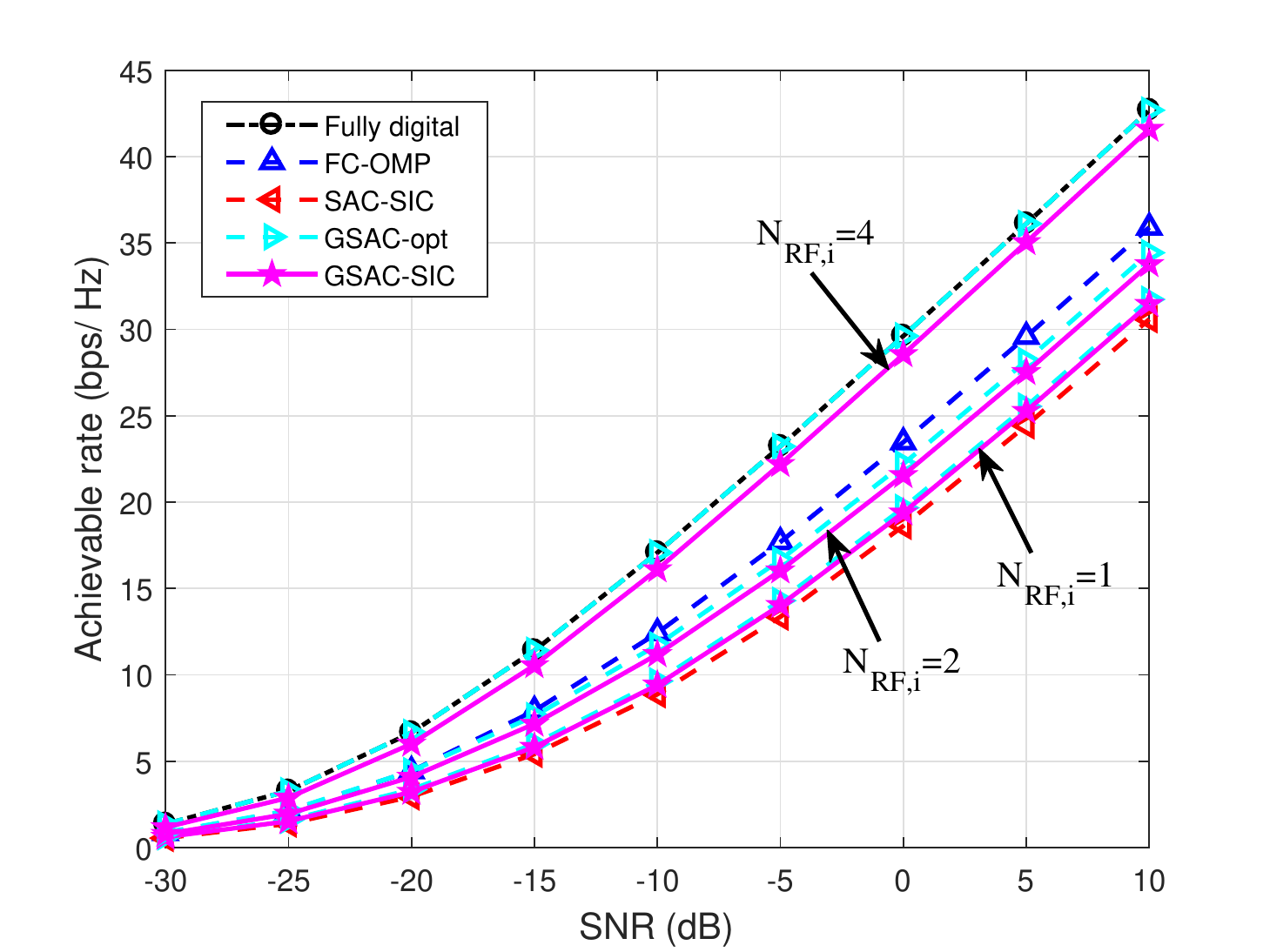}
  \caption{Achievable rate vs SNR with $N_{\rm t} = 144$, $N_{\rm r} = 36$ and ${N_{{\rm{RF}}}^{\rm t}} = 4$.}
  \label{SE vs SNR}
\end{figure}
Fig. \ref{SE vs SNR} shows the achievable rates of the proposed GSAC-SIC scheme, GSAC-opt scheme, FC-OMP scheme and SAC-SIC scheme, where we have $N_{\rm t} = 144$, $N_{\rm r} = 36$ , ${N_{{\rm{RF}}}^{\rm t}} = 4$ and we assume $N_{{\rm{RF}},1} =, ..., = N_{{\rm{RF}}, N_{\rm sub}} = N_{{\rm{RF}},i}$. When $N_{{\rm{RF}},i} = 1$, the GSAC architecture represents the traditional SAC architecture. Observe in Fig. \ref{SE vs SNR} that the achievable rate of the proposed GSAC-SIC scheme is similar to (little higher than ) that of the SAC-SIC scheme. When $N_{{\rm{RF}},i} = 4$, which means that there is only a single sub-array, the GSAC architecture now becomes the FC architecture. It can be observed that the unconstrained TPC scheme and the GSAC-SIC scheme achieve a similar rate as the fully digital scheme. When $N_{{\rm{RF}},i} = 2$, we can observe that the proposed GSAC-SIC scheme still achieves a similar rate as the FC-OMP scheme.

\begin{figure}[t]
  \centering
  \includegraphics[scale=0.5]{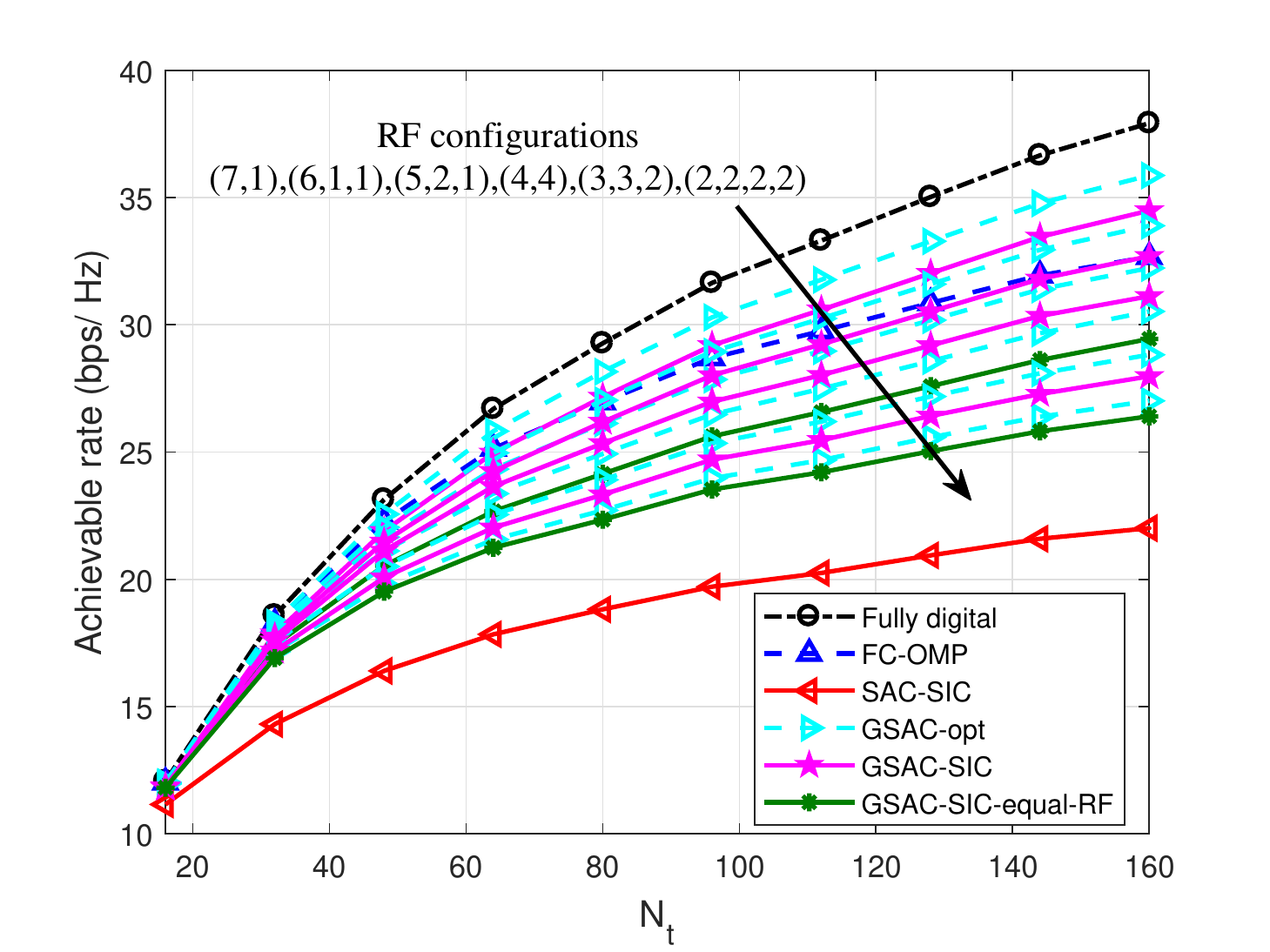}%
  \caption{Achievable rate vs $N_{\rm t}$. with $N_{\rm r} = 36$ and $N_{\rm s} = {N_{{\rm{RF}}}^{\rm t}} = 8$.}
  \label{Achievable rate_Nt}
\end{figure}

\begin{figure}[t]
  \centering
  \includegraphics[scale=0.5]{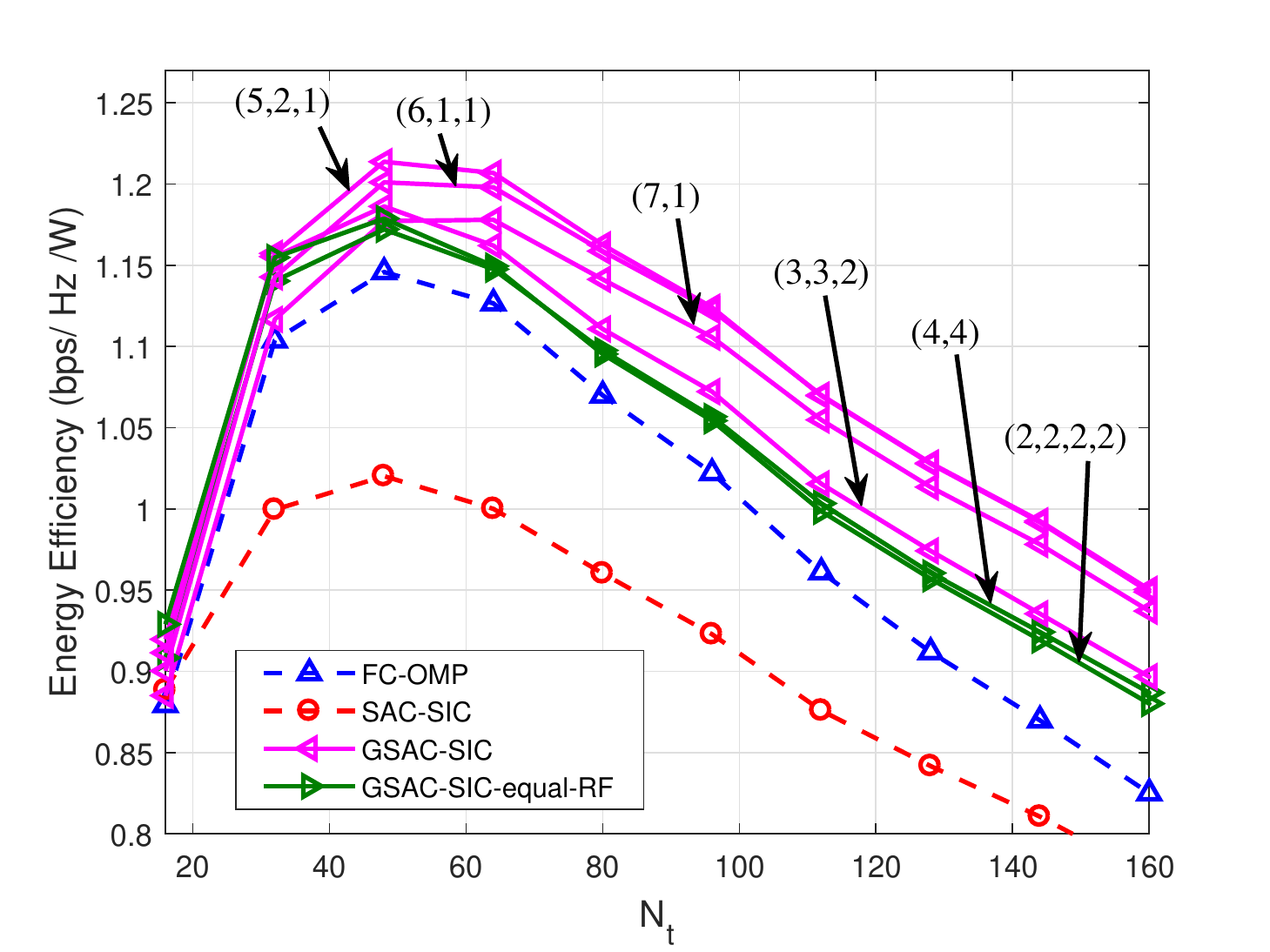}%
  \caption{EE vs $N_{\rm t}$ with $N_{\rm r} = 36$ and $N_{\rm s} = {N_{{\rm{RF}}}^{\rm t}} = 8$.}
  \label{Energy efficiency_Nt}
\end{figure}

\begin{figure}[t]
\centering
\subfigure[Achievable rate vs $N_{\rm sub}$ .]{
\begin{minipage}{0.45\textwidth}
\centering
\includegraphics[scale=0.5]{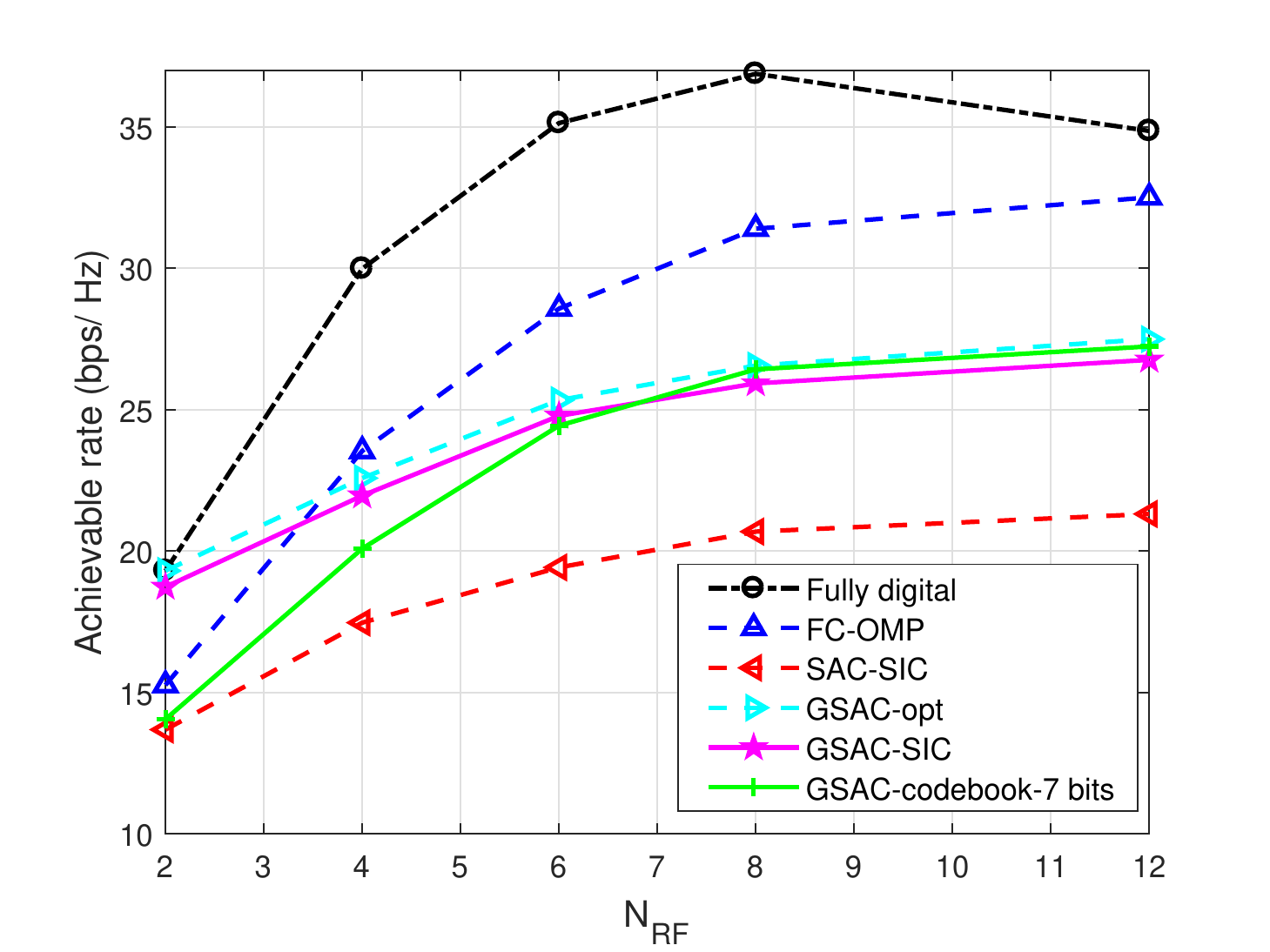}
\end{minipage}
}
\subfigure[Energy efficiency vs $N_{\rm sub}$.]{
\begin{minipage}{0.45\textwidth}
\centering
\includegraphics[scale=0.5]{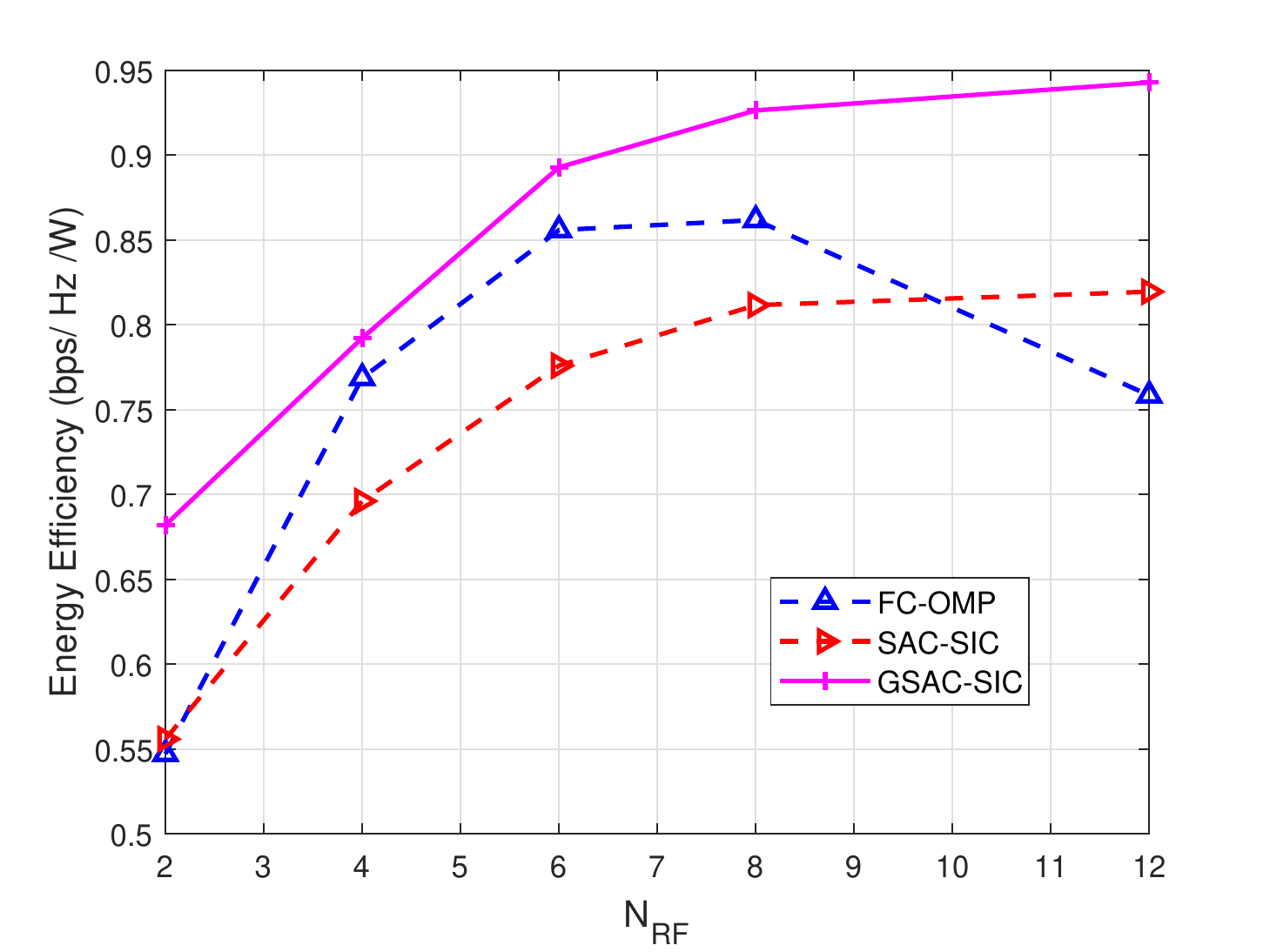}
\end{minipage}
}
\caption{Achievable rate and energy efficiency with different $N_{\rm sub}$, when $N_{\rm t} = 144$, $N_{\rm r} = 36$ and ${N_{{\rm{RF}},i}} = 2$.}
\label{fig3}
\end{figure}

\begin{figure}[t]
\centering
\subfigure[Achievable rate vs ${N_{{\rm{RF}},i}}$.]{
\begin{minipage}{0.45\textwidth}
\centering
\includegraphics[scale=0.5]{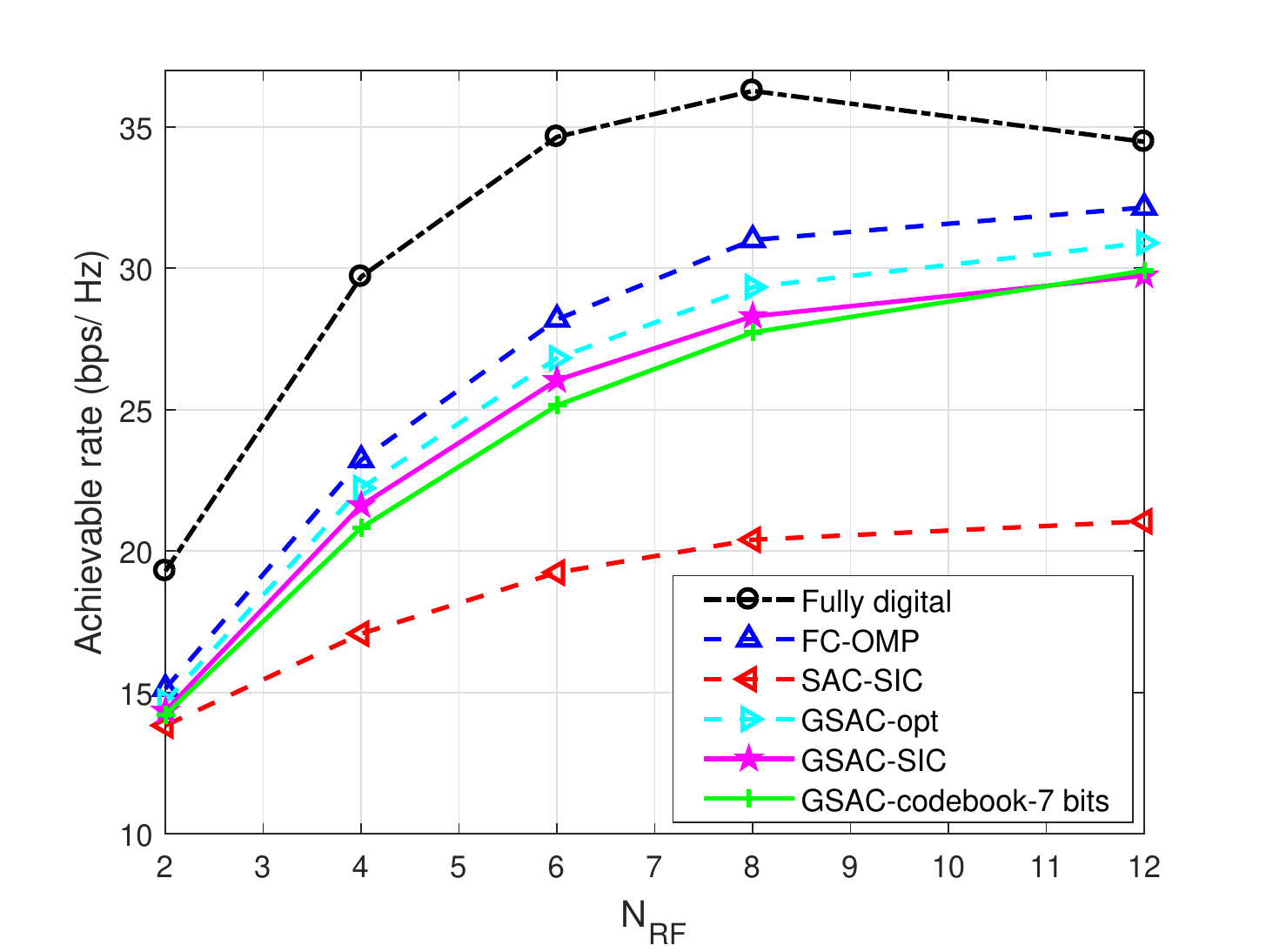}
\end{minipage}
}
\subfigure[Energy efficiency vs ${N_{{\rm{RF}},i}}$.]{
\begin{minipage}{0.45\textwidth}
\centering
\includegraphics[scale=0.5]{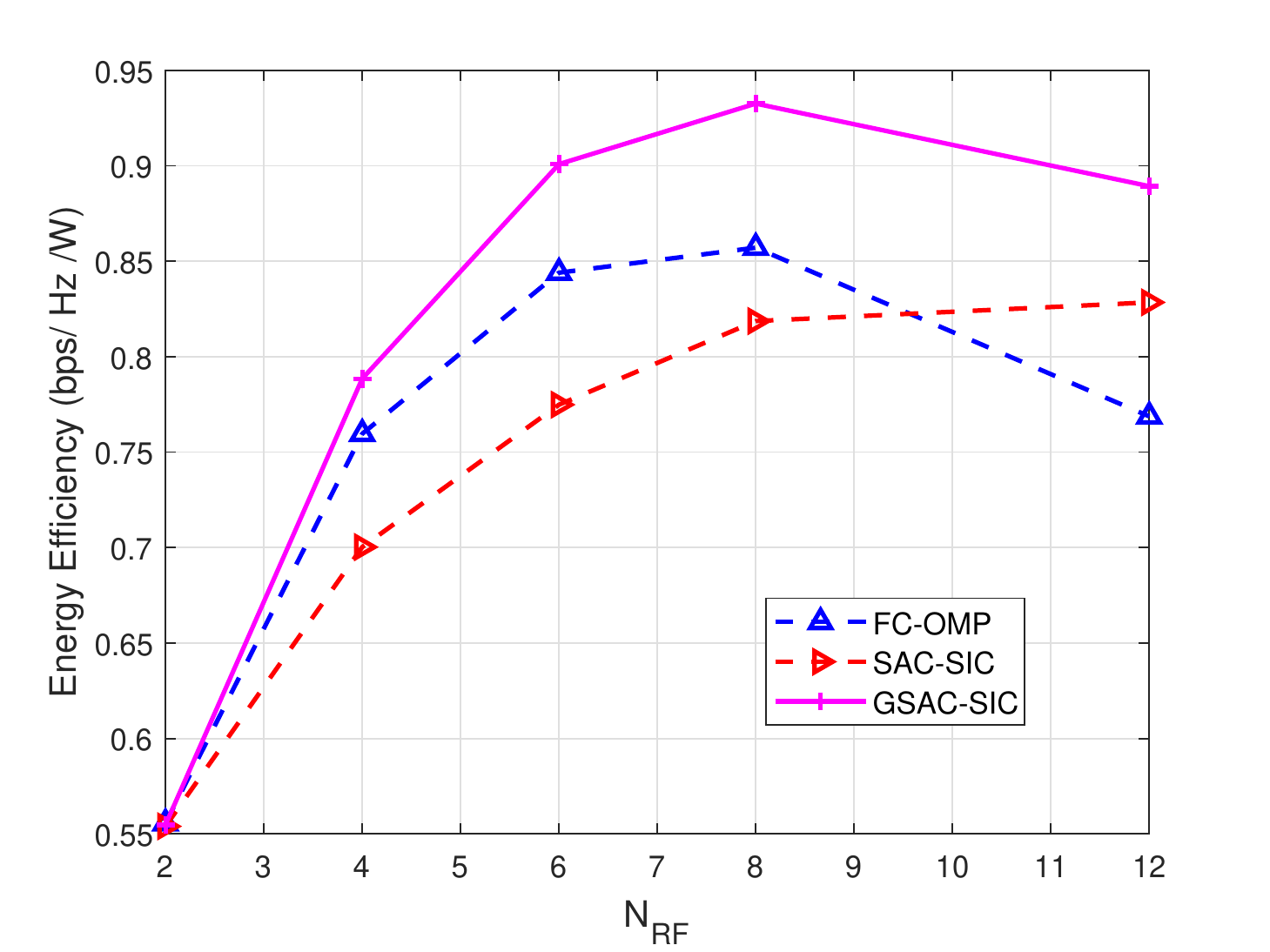}
\end{minipage}
}
\caption{Achievable rate and energy efficiency with different ${N_{{\rm{RF}},i}}$, when $N_{\rm t} = 144$, $N_{\rm r} = 36$ and  $N_{\rm sub} = 2$.}
\label{fig4}
\end{figure}
Fig. \ref{Achievable rate_Nt} shows the achievable rate for different numbers of antennas at the transmitter, when $N_{\rm r} = 36$,  $N_{\rm s} = {N_{{\rm{RF}}}^{\rm t}} = 8$. After preliminary screening by \textbf{Algorithm 2}, we select some relatively better configurations, which are (7, 1), (6, 1, 1), (5, 2, 1), (4, 4), (3, 3, 2) and (2, 2, 2, 2). For example, the mapping (5, 2, 1) means that there are 3 sub-arrays and the numbers of RF chains connected to the sub-arrays are 5, 2 and 1, respectively. At the same time, according to (\ref{4.42}), the corresponding numbers of antennas connected to sub-arrays are 5, 2 and 1 multiplied by ${N_{{\rm{t}}}}/{N_{{\rm{RF}}}}$. It can be observed that the proposed GSAC-SIC scheme achieves a similar rate to that of the corresponding optimal unconstrained TPC scheme in (\ref{3.26}) and we have $R_{(7, 1)} > R_{(6, 1, 1)} > R_{(5, 2, 1)} >  R_{(4, 4)} > R_{(3, 3, 2)} > R_{(2, 2, 2, 2)} > R_{\rm SAC-SIC}$. Specifically, we could observe that the achievable rate of the proposed GSAC-SIC scheme under the configuration (6, 1, 1) is similar to that of the FC-OMP scheme in the FC architecture and the proposed GSAC-SIC scheme under the configuration (7, 1) obtains better achievable rate than the FC-OMP scheme. The above results are due to the analog TPC vectors of the OMP scheme is selected from a predefined dictionary matrix, but the analog TPC vectors of the proposed GSAC-SIC scheme precisely extract phases of elements in the corresponding optimal unconstraint matrix.

Fig. \ref{Energy efficiency_Nt} compares the EE of different schemes for different numbers of antennas at the transmitter. The system parameters are the same as in Fig. \ref{Achievable rate_Nt}. It may be observed that the proposed GSAC-SIC scheme outperforms both the FC and the SAC architectures in terms of EE. As for the peak values of EEs, the configuration (5, 2, 1) is the best. Additionally, the configurations (6, 1, 1) and (5, 2, 1) are better than (4, 4) and (2, 2, 2, 2), where the EEs of the configurations (4, 4) and (2, 2, 2, 2) are nearly the same. Furthermore, it may also be observed that the EEs of all schemes increase first and then decrease as the number of antennas increases. For the system parameters considered (i.e., $N_{\rm r} = 36$, $N_{\rm s} = {N_{{\rm{RF}}}^{\rm t}} = 8$), the EE is highest when the number of antennas is about 48.

Fig. \ref{fig3} and Fig. \ref{fig4} compare the achievable rate and EE with different $N_{\rm sub}$ and ${N_{{\rm{RF}},i}}$, respectively, where $N_{\rm t} = 144$, $N_{\rm r} = 36$, $N_{\rm s} = {N_{{\rm{RF}}}^{\rm t}}$.  We could observe that the achievable rates of the GSAC-SIC scheme are always larger than the SAC-SIC scheme and the EE of the GSAC-SIC scheme is the best compared with the FC-OMP scheme in the FC architecture and the SAC-SIC scheme in the SAC architecture. In addition, in Fig. \ref{fig3} (a), when ${N_{{\rm{RF}}}^{\rm t}} = 2$, i.e., $N_{\rm sub} = 1$ and ${N_{{\rm{RF}},i}} = 2$, the GSAC-SIC scheme achieves similar achievable rate as the fully digital scheme. In Fig. \ref{fig4} (a), when ${N_{{\rm{RF}}}^{\rm t}}= 2$, i.e., $N_{\rm sub} = 2$ and ${N_{{\rm{RF}},i}} = 1$, the GSAC-SIC scheme achieves similar achievable rate as the SAC-SIC scheme. Both the above observations verify the near-optimal performance of the GSAC-SIC scheme. Moreover, as for the proposed GSAC-codebook scheme, the quantization bits are set to be 7 in both Fig. \ref{fig3} and Fig. \ref{fig4}. We can observe that the achievable rate of the GSAC-codebook scheme is better than the SAC-SIC scheme and is similar as the GSAC-SIC scheme, which verify the effectiveness of the quantization in \textbf{Algorithm 3}. Finally, it can be also observed that, when applying more numbers of RF chains, the achievable rates do not always increase. This is because $N_{\rm s}= N_{\rm RF}$ with equal power allocation is assumed and the total transmit power is constrained. When transmitting more data streams, the average power allocated to single data stream decreases.

\begin{figure}[t]
  \centering
  \includegraphics[scale=0.5]{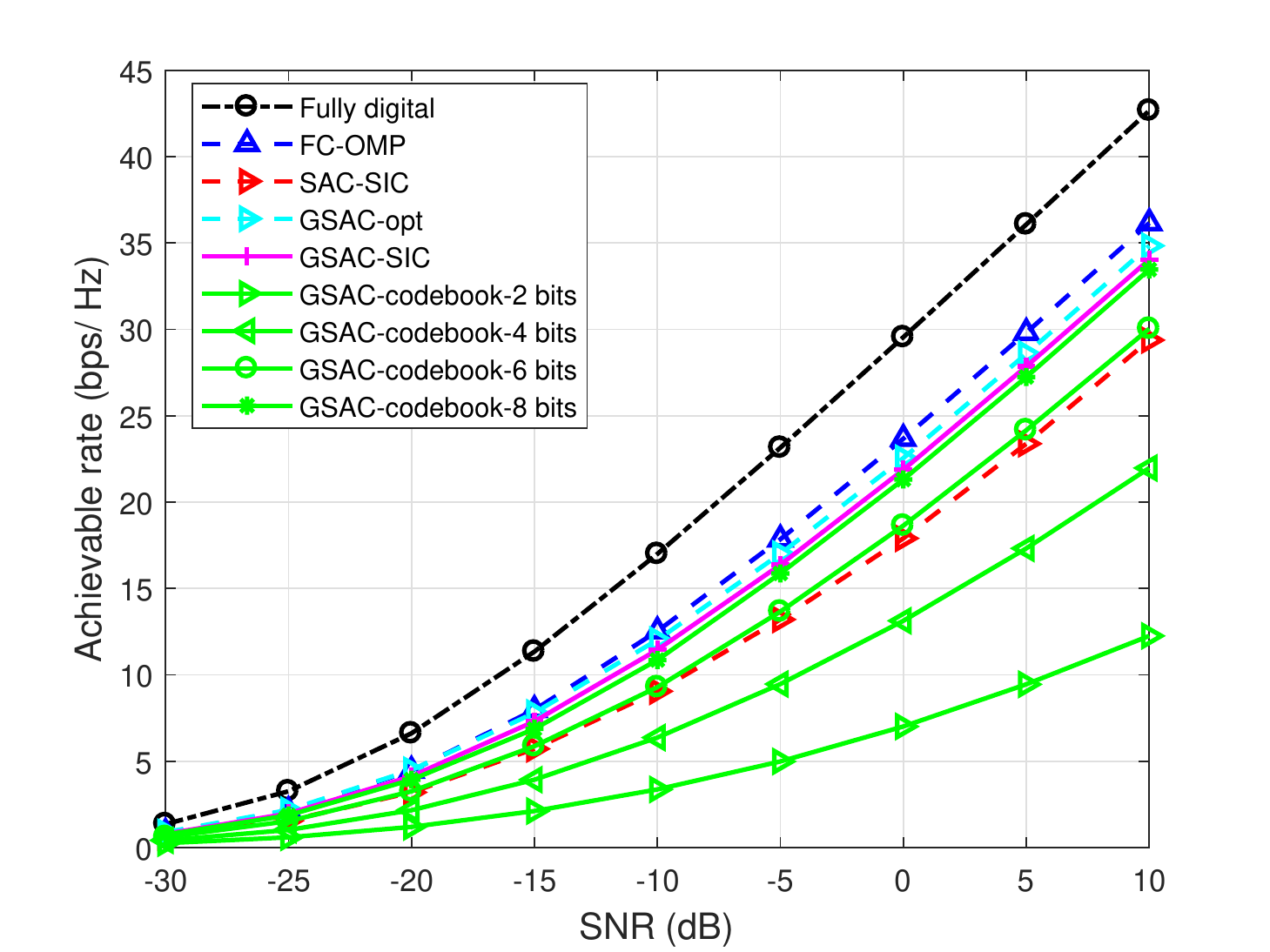}%
  \caption{Achievable rate for different quantization bits with $N_{\rm t} = 144$, $N_{\rm r} = 36$, $N_{\rm s} = {N_{{\rm{RF}}}^{\rm t}} = 4$ and ${N_{{\rm{RF}},i}} = 2$.}
  \label{Achievable_rate_codebook4}
\end{figure}

\begin{figure}[t]
  \centering
  \includegraphics[scale=0.5]{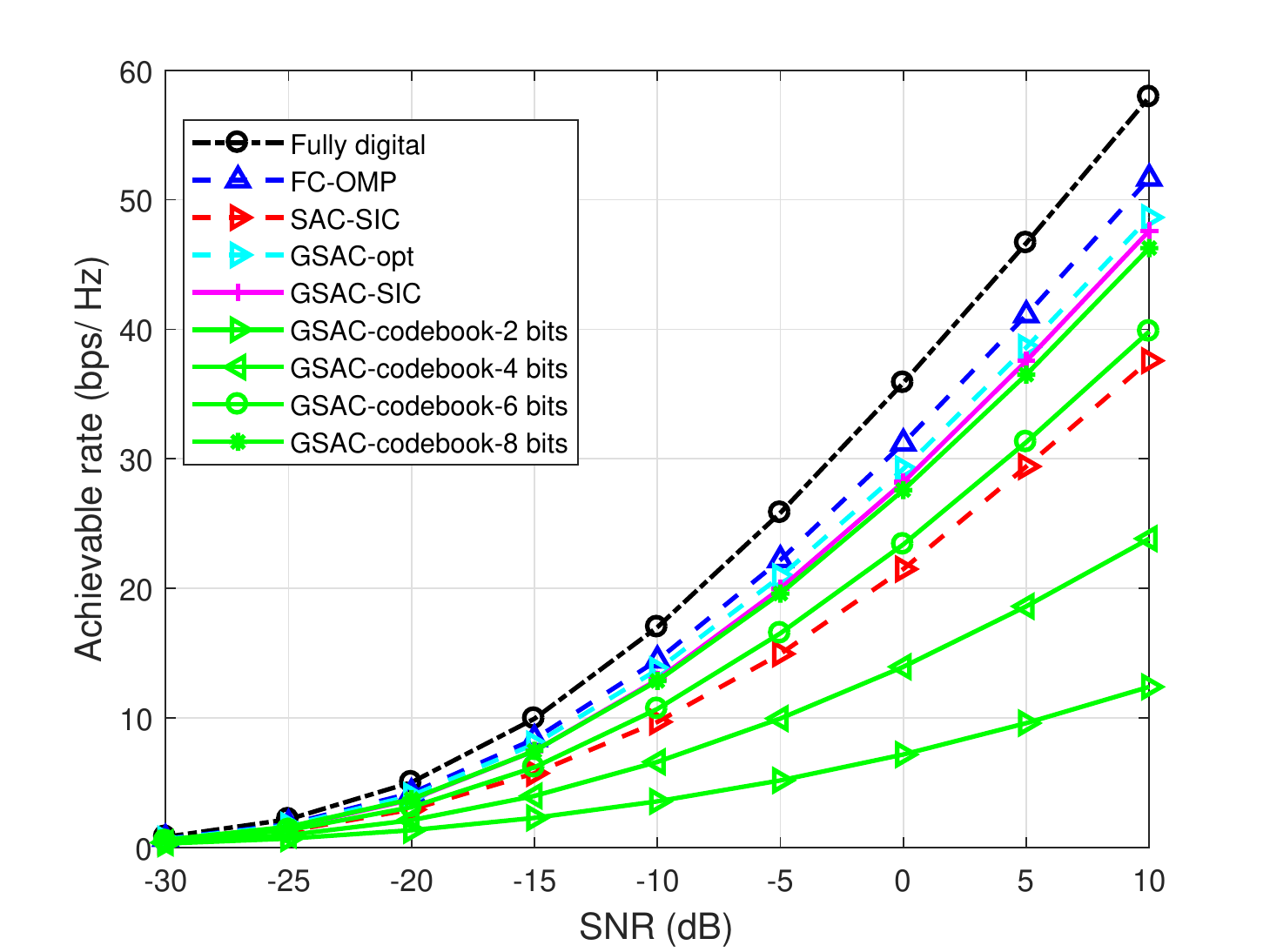}%
  \caption{Achievable rate for different quantization bits with $N_{\rm t} = 144$, $N_{\rm r} = 36$, $N_{\rm s} = {N_{{\rm{RF}}}^{\rm t}}= 8$, and ${N_{{\rm{RF}},i}} = 4$.}
  \label{Achievable rate_codebook8}
\end{figure}
Fig. \ref{Achievable_rate_codebook4} and Fig. \ref{Achievable rate_codebook8} show the effect of the quantization bit on the achievable rate of the proposed GSAC-codebook scheme under ${N_{{\rm{RF}}}^{\rm t}} = 4$ and ${N_{{\rm{RF}}}^{\rm t}} = 8$, respectively. For both figures, $N_{\rm sub} = 2$ and the quantization bit $b$ varies from 6 to 8. First, it can be seen that the proposed GSAC-SIC algorithm achieves similar achievable rate as the FC-OMP scheme for the given RF configurations, i.e., (2, 2) and (4, 4). Then, we could observe that even for $b=6$ bits, the achievable rate of the proposed GSAC-codebook scheme is better than the SAC-SIC scheme. Finally, Fig. \ref{Achievable_rate_codebook4} and Fig. \ref{Achievable rate_codebook8} also show that for the situation when $b=8$ bits, the proposed GSAC-codebook scheme achieves similar achievable rate as the corresponding optimal unconstrained TPC scheme.

\section{Conclusions}
In this paper, a GSAC architecture was considered for improving the EE of hybrid TPC in millimeter-wave massive MIMO systems relying on arbitrary RF and antenna configurations. For any given RF and antenna configuration, a SIC based algorithm with near-optimal achievable rate was firstly proposed. In order to find the configuration having the best energy efficiency, an exhaustive yet modest-complexity search scheme was proposed. Moreover, to facilitate using a limited feedback in a practical millimeter wave MIMO system, a beamsteering codebook based hybrid TPC scheme was proposed for the GSAC architecture. Our simulation results verified that the proposed scheme achieves a similar rate to the corresponding optimal unconstrained TPC scheme and attains the best energy-efficiency among the schemes investigated.

\begin{IEEEbiography}[{\includegraphics[width=1in,height=1.25in]{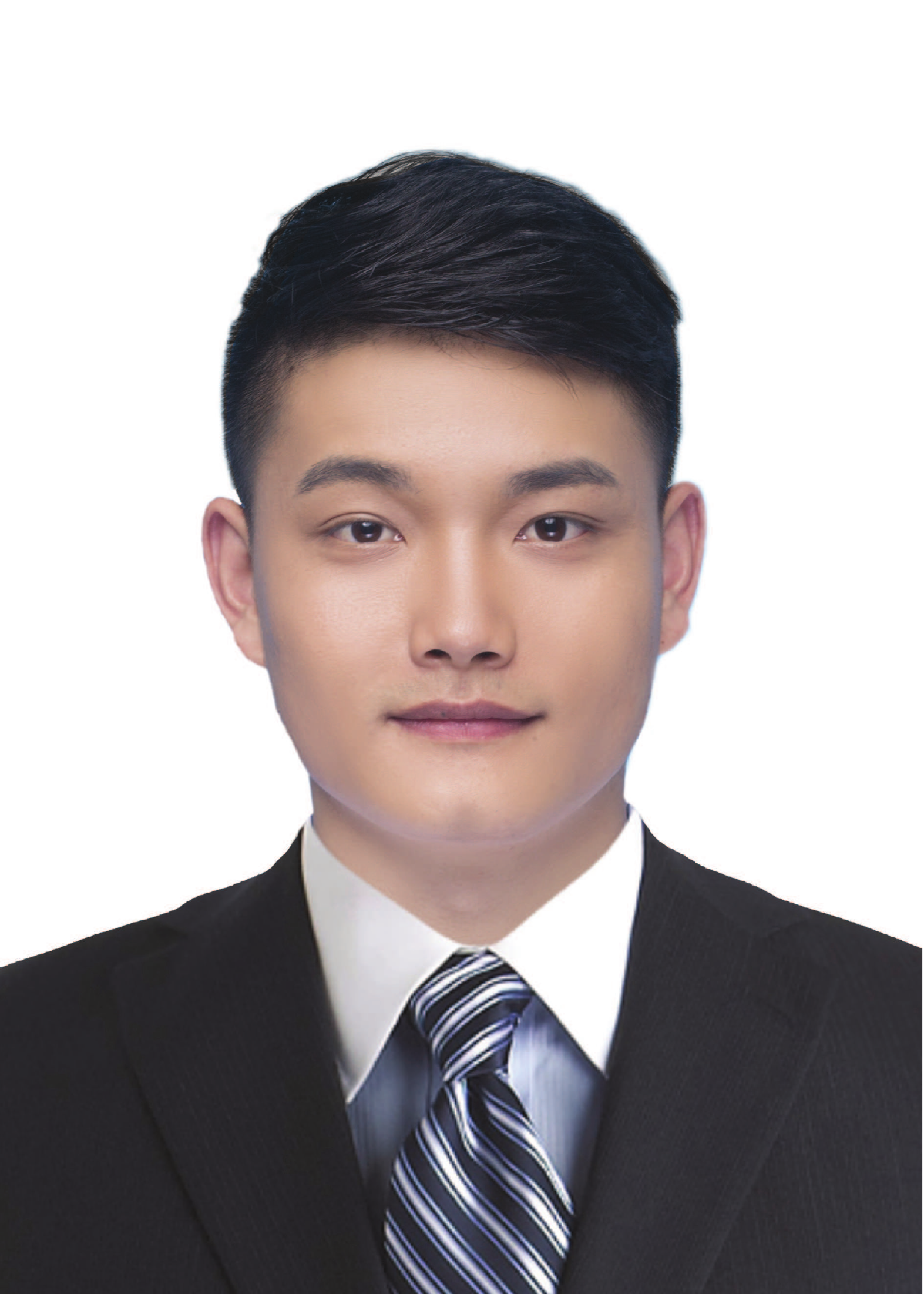}}]
{\bf Yun Chen} received the B.S. degree from Huazhong University of Science and Technology, Wuhan, P. R. China, in 2016, where he is currently pursuing the Ph.D degree with Wuhan National Laboratory for Optoelectronics and School of Electronic Information and Communications. Since 2018, he has been a Visiting Student with the School of Electronics and Computer Science, University of Southampton, U.K. His current research interests include millimeter wave communications and machine learning.
\end{IEEEbiography}

\begin{IEEEbiography}[{\includegraphics[width=1in,height=1.25in]{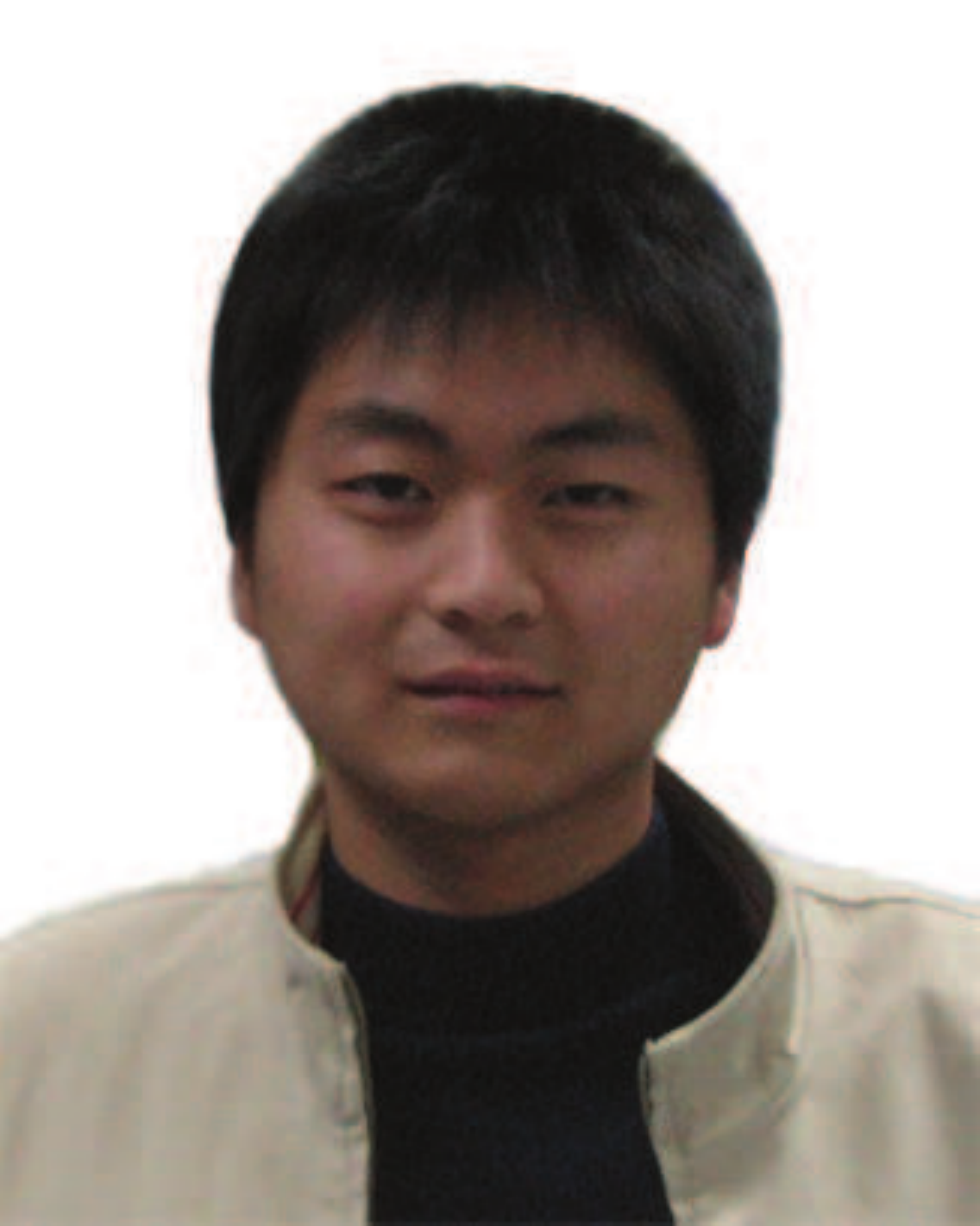}}]
{\bf Da Chen} received the B.S. and Ph.D. degrees from  Huazhong University of Science and Technology, Wuhan, P. R. China, in 2009 and 2015, respectively. From Sep. 2012 to Aug. 2013, he was a visiting scholar at Northwestern University, USA. From Sep. 2013 to Sep. 2014, he was a visiting scholar at University of Delaware, USA. He is currently an Assistant Professor with the School of Electronics Information and Communications, Huazhong University of Science and Technology, Wuhan, P. R. China. He is serving as an Associate Editor for China Communications. His current research interests include various areas in wireless communications, such as OFDM and FBMC systems.
\end{IEEEbiography}

\begin{IEEEbiography}[{\includegraphics[width=1in,height=1.25in]{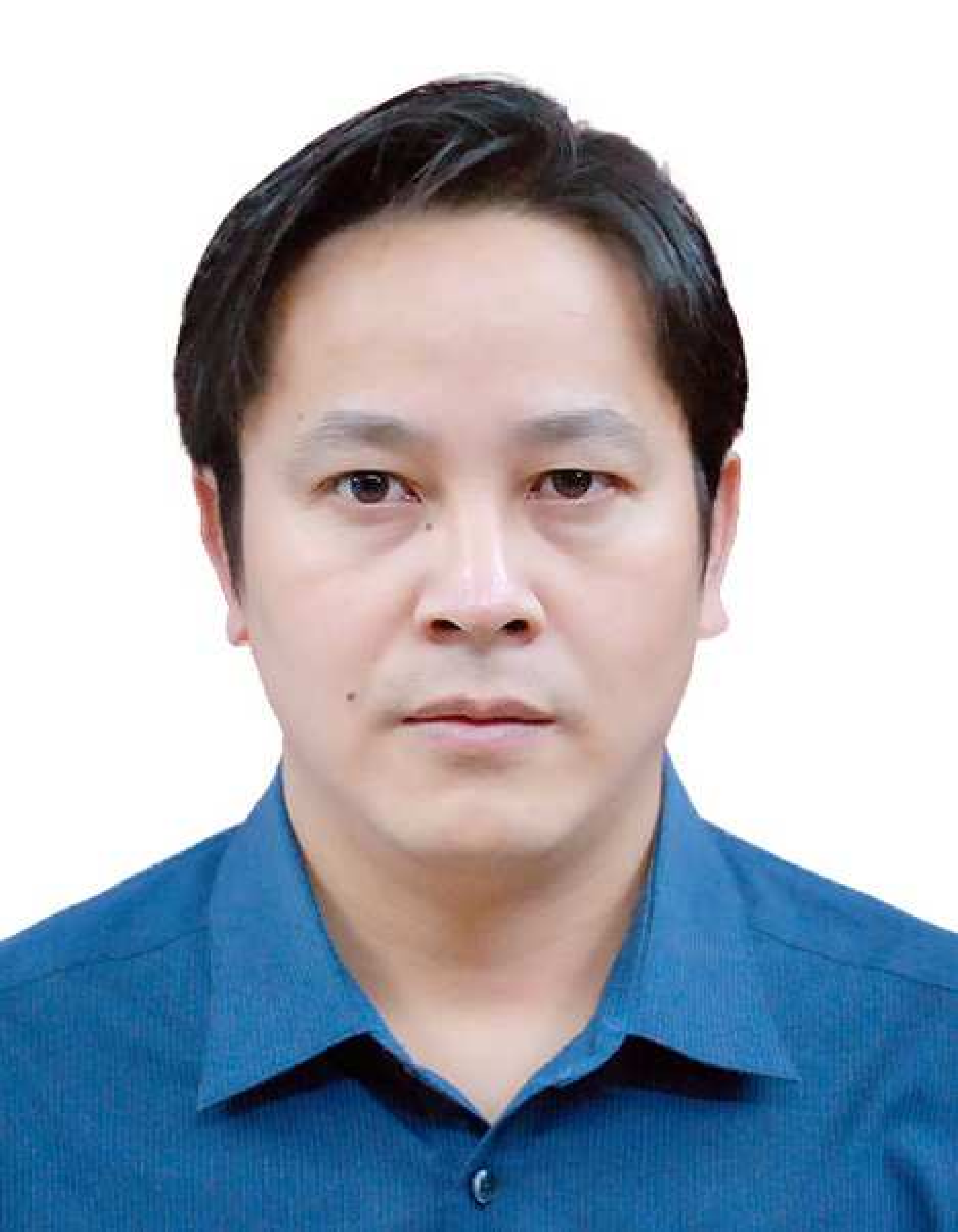}}]
{\bf Tao Jiang} (M'06-SM'10-F'19) is currently a Distinguished Professor in the Wuhan National Laboratory for Optoelectronics and School of Electronics Information and Communications, Huazhong University of Science and Technology, Wuhan, P. R. China. He received Ph.D. degree in information and communication engineering from Huazhong University of Science and Technology, Wuhan, P. R. China, in April 2004. From Aug. 2004 to Dec. 2007, he worked in some universities, such as Brunel University and University of Michigan-Dearborn, respectively. He has authored or co-authored more 300 technical papers in major journals and conferences and 9 books/chapters in the areas of communications and networks. He served or is serving as symposium technical program committee membership of some major IEEE conferences, including INFOCOM, GLOBECOM, and ICC, etc.. He was invited to serve as TPC Symposium Chair for the IEEE GLOBECOM 2013, IEEEE WCNC 2013 and ICCC 2013. He is served or serving as associate editor of some technical journals in communications, including in IEEE Network, IEEE Transactions on Signal Processing, IEEE Communications Surveys and Tutorials, IEEE Transactions on Vehicular Technology, IEEE Internet of Things Journal, and he is the associate editor-in-chief of China Communications, etc..
\end{IEEEbiography}

\begin{IEEEbiography}[{\includegraphics[width=1in,height=1.25in]{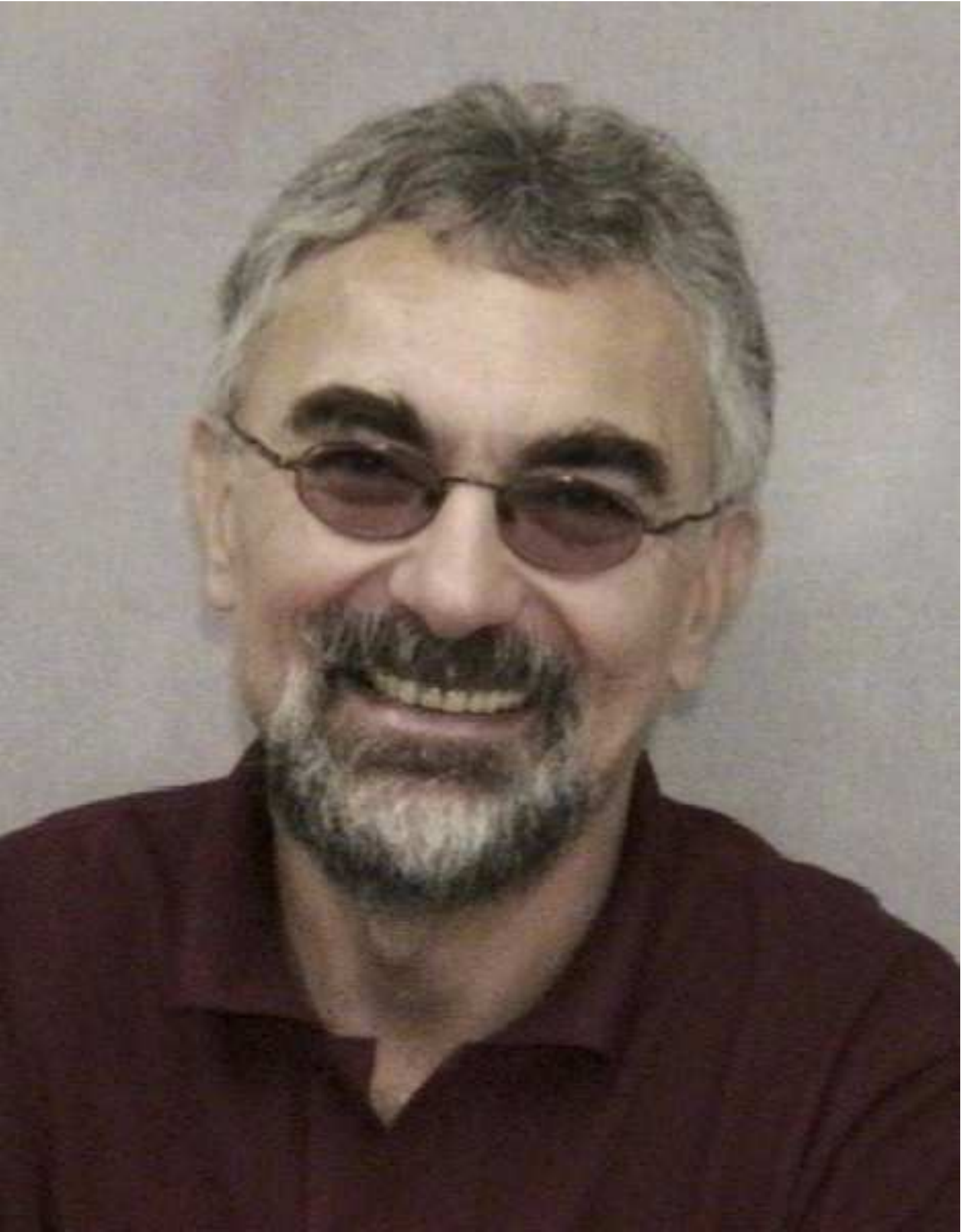}}]
{\bf Lajos Hanzo} (M'91-SM'92-F'04) FREng, FIET, Fellow of EURASIP, received his 5-year degree in electronics in 1976 and his doctorate in 1983 from the Technical University of Budapest.  In 2009 he was awarded an honorary doctorate by the Technical University of Budapest and in 2015 by the University of Edinburgh.  In 2016 he was admitted to the Hungarian Academy of Science. During his 40-year career in telecommunications he has held various research and academic posts in Hungary, Germany and the UK. Since 1986 he has been with the School of Electronics and Computer Science, University of Southampton, UK, where he holds the chair in telecommunications.  He has successfully supervised 112 PhD students, co-authored 18 John Wiley/IEEE Press books on mobile radio communications totalling in excess of 10 000 pages, published 1760 research contributions at IEEE Xplore, acted both as TPC and General Chair of IEEE conferences, presented keynote lectures and has been awarded a number of distinctions. Currently he is directing a 60-strong academic research team, working on a range of research projects in the field of wireless multimedia communications sponsored by industry, the Engineering and Physical Sciences Research Council (EPSRC) UK, the European Research Council's Advanced Fellow Grant and the Royal Society's Wolfson Research Merit Award.  He is an enthusiastic supporter of industrial and academic liaison and he offers a range of industrial courses.  He is also a Governor of the IEEE ComSoc and VTS.  During 2008 - 2012 he was the Editor-in-Chief of the IEEE Press and a Chaired Professor also at Tsinghua University, Beijing. For further information on research in progress and associated publications please refer to http://www-mobile.ecs.soton.ac.uk.
\end{IEEEbiography}

\end{document}